\documentclass
[superscriptaddress,secnumarabic,amssymb,amsmath,nobibnotes,aps,prd,nofootinbib]{revtex4}%

\usepackage[T1]{fontenc}
\usepackage{tikz-cd}
\usepackage{hyperref}
\usepackage{amsmath}
\usepackage{amsthm}
\usepackage{amsfonts}
\usepackage{xcolor}
\usepackage{amssymb}
\usepackage{stmaryrd}
\usepackage{soul}
\usepackage{graphicx,subcaption}

\usepackage[toc,page]{appendix}

\allowdisplaybreaks
\def\be{\begin{equation}}
\def\ee{\end{equation}}
\def\bea{\begin{eqnarray}}
\def\eea{\end{eqnarray}}

\def\hb{\hat{B}}
\def\hc{\hat{C}}

\def\hf{\hat{F}}
\def\hg{\hat{g}}
\def\hp{\hat{\phi}}
\def\hx{\hat{X}}

\begin{document}

\def\ra{\longrightarrow}
\def\ci{\mathcal I}
\def\ca{{\mathcal A}}
\def\cb{{\mathcal B}}
\def\cc{{\mathcal C}}
\def\cd{{\mathcal D}}
\def\ch{{\mathcal H}}%
\def\cm{{\mathcal M}}
\def\cv{{\mathcal V}}
\def\cw{{\mathcal W}}
\def\cR{{\mathcal R}}
\def\hr{{\hat R}}
\def\cH{{\mathcal H}}
\def\fH{{\mathfrak H}}
\def\cT{{\mathcal T}}
\def\dph{{\dot{\Phi}}}
\def\aa{{\mathfrak a}}
\def\xx{{\mathfrak y}}
\def\tT{{\mathfrak t}}
 

\title{Disformal transformations in a Palatini extension of Horndeski's gravity}

\author{Aleksander Kozak}
	\email{aleksander.kozak2@uwr.edu.pl}
\affiliation{ Departamento de Matem\'{a}ticas, Universidad Católica del Norte, Avda. Angamos 0610, Casilla 1280, Antofagasta, Chile \\ 
University of Wrocław, Institute of Theoretical Physics, pl. Maxa Borna 9, 50-206 Wroclaw, Poland}


\begin{abstract}
In this paper, we extend Horndeski's theory into the Palatini approach, assuming that the metric tensor and the (symmetric) connection are \textit{a priori} independent objects. We introduce an additional transformation of the connection and write down the action functional being form-invariant under both the disformal transformation of the metric and the new transformation of the connection. We show that such a theory reduces on-shell to a metric subclass of Horndeski's gravity called kinetic gravity braiding. We also introduce an invariant metric and connection, and demonstrate that quantities defined in such a way lead to a metric theory. In the second part of the paper, we consider a simple cosmological model within the theory and explore its potential links with $ k$-essence-type theories, with a non-trivial coupling between the scalar field and the matter part of the action in the Einstein frame. We show that there exists a model that reproduce late-time cosmic acceleration, approaching asymptotically the de Sitter phase, motivating further study of the theories.
\end{abstract}



\maketitle
\section{Introduction}
Scalar-tensor (ST) theories of gravity provide possibly one of the most straightforward ways of modifying Einstein's General Relativity (GR), leading at the same time to a rich phenomenology introduced by a dynamical scalar field, coupled non-minimally to the gravitational part of the action functional \cite{capozziello:2011, fujii:2004}. ST theories have been applied in various contexts to tackle the problems faced by GR, such as the accelerated late-time cosmic expansion of the Universe \cite{faraoni:2004}, especially in the context of evolving dark energy (DE) called 'quintesence' \cite{caldwell:1998, boisseau:2000, bartolo:2000, fujii:2000}, somewhat relevant today due to the mounting evidence for a time-dependent cosmological constant \cite{DESI-dr2:2025a, DESI:2025b, tada:2024}. Moreover, an additional, dynamical scalar field could be employed to solve the cosmological lithium problem by modifying the rate of expansion of the Universe during the lithium synthesis  \cite{larena:2007, coc:2006}. Their theoretical motivation rests upon the fact that they can be obtained as a low-energy limit of more fundamental theories, such as string theory \cite{fujii:2004}, where the dilatonic field couples non-minimally to the curvature. 

Besides providing valuable explanations of the gravitational phenomena, ST theories must be sound also on the theoretical level. The crucial requirement is to lead to a theory free of any instabilities, potentially manifesting themselves as so-called Ostrogradski's ghosts. The ghosts are a danger that any higher-order theory is susceptible to \cite{ganz:2021}; for this reason, a general guideline for constructing alternatives to GR was to consider Lagrangians leading to second-order equations of motion. The most general scalar-tensor theory fulfilling this requirement is the well-known Horndeski's gravity \cite{horndeski:1974}. As it was discovered relatively recently, one can also obtain a perfectly healthy theory (in the sense of avoiding unwanted additional degrees of freedom related to the scalar field) by considering models with a degenerate kinetic matrix, called Degenerate Higher-Order Scalar-Tensor (DHOST) theories \cite{langlois:2017, achour:2016}. However, just like in the case of Horndeski's gravity, the speed of gravitational waves \cite{baker:2017} and the requirement that the waves do not decay into DE \cite{creminelli:2018} impose serious constraints on the possible forms of the action functional (in the case of Horndeski's gravity, the resulting theory, dubbed 'kinetic gravity braiding' (KGB), features only the term $G(\phi, X)\Box \phi$, with $X = (\partial\phi)^2$, in addition to the curvature and a general function $F(\phi, X)$ in the gravitational part of the action functional \cite{quartin:2023}). The surviving theory still features the Vainshtein mechanism, allowing one to hide the effect of the dynamical scalar field at the scale of the Solar System, but at the same time, it requires fine-tuning \cite{hirano:2021}, unlike in the general DHOST theories, where the screening mechanism outside of massive bodies appears somewhat naturally \cite{kobayashi:2015}.

So far, both Horndeski's theory and its generalizations have been considered mainly in the metric approach, i.e., in a framework where the dynamical quantities are the metric tensor and the scalar field. Alternatively, one could try to extend the analysis to the metric-affine (MA) approach, where one decouples the metric structure of spacetime from the affine structure, treating the connection as an independent field. The motivation for such a procedure is twofold: firstly, one can argue that the MA approach is more fundamental, as it does not impose any relation between the metric and the connection a priori; secondly, the MA paradigm might increase the freedom of choosing scalar field functions entering the action that are in agreement with the requirements outlined above. Some attention has been dedicated to Palatini (i.e., torsionless version of the MA gravity with the connection not entering the matter part of the action) ST theory in the Wagoner parametrization \cite{wagoner:1970}: the action functional was extended by introducing additional terms constructed from non-metricity \cite{kozak:2019}, and the theory itself was analyzed in particular in the context of cosmic inflation \cite{gialamas:2023, racioppi:2017, racioppi:2018, jarv:2018}. Moreover, some effort has been dedicated to constructing an MA version of Horndeski’s theory \cite{helpin:2020a, helpin:2020b, dong:2023}. It has
been demonstrated that specific subclasses of the Palatini-Horndeski’s theory reproduce purely metric
theory since the connection is an auxiliary field. In such cases, additional scalar-tensor interactions might
lead to ghosts \cite{helpin:2020b}. However, it remains to be seen whether introducing extra terms resulting from
adopting the MA approach can always render the theory stable. Partial results exist: it was demonstrated that some MA Horndeski’s theories can be ghost-free when
terms proportional to non-metricity are introduced \cite{helpin:2020b}. Imposing the constraint related to the speed of gravitational waves on the Palatini-Horndeski's theory limits the possible space of functions to the Palatini KGB parameterization, which, after integrating out the connection, turns out to be of the same form as the metric KGB model \cite{helpin:2020b, dong:2023}. 

Horndeski's theory with a general auxiliary connection as analyzed, for example, in \cite{helpin:2020a}, is of particular interest here because it introduces no additional degree of freedom; otherwise, it might be problematic due to potential instabilities \cite{dong:2023}. Such a connection can always be integrated out, thus leading to an on-shell-equivalent metric theory with well-established stability properties and the correct speed of gravitational waves. A simple example includes Palatini $f(R)$ gravity, which can be viewed as an ST theory with a non-dynamical scalar field and an auxiliary connection \cite{vitagliano:2011, olmo:2011, camera:2023, li:2007, stachowski:2017}, effectively leading to GR with a modified source. 

Of special importance to ST theories are conformal and disformal transformations of the metric tensor. It was established that the class of Horndeski's theory remains stable under the transformation of the form $g_{\mu\nu}\rightarrow C(\phi)g_{\mu\nu} + D(\phi)\phi_\mu\phi_\nu$, with $\phi_\mu=\partial_\mu\phi$, whereas the DHOST theories preserve their form under a more general transformation $g_{\mu\nu}\rightarrow C(\phi, X)g_{\mu\nu} + D(\phi, X)\phi_\mu\phi_\nu$ \cite{langlois:2017}. The idea of applying the disformal transformation of the metric in the context of the Palatini gravity has been explored by some authors \cite{gialamas:2020, galtsov:2019, nezhad:2024, annala:2021}: in general, such a transformation proves useful if the theory depends on the symmetric part of the Ricci tensor. Disformal transformations have also been analyzed in the context of the teleparallel gravity \cite{golovnev:2020, hohmann:2019}. Applying such a transformation can potentially allow one to change the parametrization of the theory in which the model turns out to be effectively metric. The behaviour of the Palatini extension of Horndeski's gravity under such transformations remains, however, unexplored; one can hope, though, that, upon transforming the metric, it will be possible to establish an equivalence class of a given theory that is in agreement with both the theoretical and observational requirements. Moreover, the Palatini approach provides the freedom to transform the connection in a way independent of the metric tensor, so that one can identify models that are mathematically equivalent also by means of such an independent transformation. An example of independent transformations of the metric tensor and the affine connection has been provided in Ref. \cite{kozak:2019}. 

The main purpose of this paper is to construct and investigate a Palatini extension of the surviving Horndeski's gravity that, at the same time, would be closed under the disformal transformation of the metric and a transformation of the connection, akin to the generalised almost-geodesic mapping introduced in Ref. \cite{kozak:2019}. The proposed theory will turn out to encompass a variety of models analysed previously in the literature, introducing novel terms into the action functional. 

The paper is organized as follows: first, in Section \ref{sec:2}, we will introduce the disformal transformation of the metric tensor and an independent transformation of the connection, and discuss their properties. In Section \ref{sec:3}, we will write down the action functional for the theory that is form-invariant under the proposed transformations. Transformation formulae for the scalar field functions present in the action will also be discussed. In the same section, we will analyze how the transformations should be correctly composed if one performs two consecutive disformal frame changes. We will also explore relations between the proposed theory and other models investigated so far in the literature. Lastly, the invariant metric and connection will be introduced. In Section \ref{sec:4}, field equations for the theory will be obtained and solved for the connection, bringing the theory to a metric, dynamically equivalent counterpart. In Section \ref{sec:5}, a simple cosmological model will be investigated with the purpose of exploring the possibility of obtaining accelerated cosmic expansion in the model, as well as illustrating how the transformations serve to simplify calculations. The paper then ends with conclusions and proposed future lines of research.

\section{Disformal transformations and the transformation of the connection}\label{sec:2}

Let us begin by introducing the notion of a general disformal transformation of the metric tensor:
\begin{equation}\label{eq:transf1}
    \bar{g}_{\mu\nu} = \alpha_1(\phi, X) g_{\mu\nu} + \alpha_2(\phi, X)\phi_\mu\phi_\nu,
\end{equation}
where the bar above the object denotes the 'old' metric defined in terms of the 'new', unbarred metric, the scalar field, and the kinetic term for the metric, $X = g^{\mu\nu}\phi_\mu\phi_\nu$. Notice that this definition uses the inverse metric $g^{\mu\nu}$ introduced in order to construct the scalar $X$. In what follows, we will assume that always $\alpha_1(\phi, X) \neq 0$. 

The inverse metric transforms according to:
\begin{equation}\label{eq:transf2}
    \bar{g}^{\mu\nu} = \beta_1(\phi, X) g^{\mu\nu} + \beta_2(\phi, X)\phi^\mu\phi^\nu,
\end{equation}
with $\phi^\mu = g^{\mu\nu}\phi_\nu$. The old metric satisfies the obvious relation $\bar{g}^{\mu\alpha} \bar{g}_{\alpha\nu} = \delta^\mu_\nu$, so that we must impose the following relations:
\begin{equation}\label{eq:inverse_met}
    \beta_1(\phi, X) = \alpha_1(\phi, X)^{-1}, \quad \beta_2(\phi, X) = \frac{-\alpha_2(\phi, X)}{\alpha_1(\phi, X)\big(\alpha_1(\phi, X)+\alpha_2(\phi, X)X\big)}.
\end{equation}
In the remaining part of the paper, we will assume that $\alpha_1(\phi, X)+\alpha_2(\phi, X)X\neq 0$. 
Forthermore, it must be possible to invert the relations \eqref{eq:transf1} and \eqref{eq:transf2} to get:
\begin{align}\label{eq:transf_met_inv}
    & g_{\mu\nu} = \bar{\alpha}_1(\phi, \bar{X})\bar{g}_{\mu\nu}+\bar{\alpha}_2(\phi, \bar{X})\phi_\mu\phi_\nu, \\
    & g^{\mu\nu} = \bar{\beta}_1(\phi, \bar{X})\bar{g}^{\mu\nu}+\bar{\beta}_2(\phi, \bar{X})\bar{\phi}^\mu\bar{\phi}^\nu,
\end{align}
with $\bar{X} = \bar{g}^{\mu\nu}\phi_\mu\phi_\nu$ and $\bar{\phi}^\mu = \bar{g}^{\mu\nu}\phi_\nu$. The relation between $\bar{X}$ and $X$ is the following:
\begin{equation}
    \bar{X} = \big(\beta_1(\phi, X)+\beta_2(\phi, X)X\big)X = \frac{X}{\alpha_1(\phi,X)+\alpha_2(\phi,X)X}.
\end{equation}
This relation must be invertible for $X$, so that the following must hold:
\begin{equation}
    \frac{\partial \bar{X}}{\partial X} = \frac{\alpha_1(\phi,X)-\alpha_{1, X}(\phi,X)X-\alpha_{2, X}(\phi,X) X^2}{(\alpha_1(\phi,X)+\alpha_2(\phi,X)X)^2}\neq 0,
\end{equation}
(where $\alpha_{i, X} = \partial_X \alpha_i$) in agreement with \cite{zumalacarregui:2014}. If this condition is satisfied, then we can establish a relation between the barred and unbarred functions $\{\alpha_i, \beta_i\}$ in the following manner:
\begin{equation}
    \bar{\alpha}_1(\phi, \bar{X}) = \beta_1(\phi, X(\bar{X})), \quad \bar{\alpha}_2(\phi, \bar{X}) = -\frac{\alpha_2(\phi, X(\bar{X}))}{\alpha_1(\phi, X(\bar{X}))},
\end{equation}
and $\bar{\beta}_1, \bar{\beta}_2$ satisfying conditions analogous to \eqref{eq:inverse_met}. Obviously, the new kinetic term will depend on the scalar field as well, but we decide to omit this dependence and write $X(\bar{X})$ instead of $X(\phi, \bar{X})$

Lastly, the determinants of the two metrics in four dimensions are related via:
\begin{equation}
    \bar{g} = \alpha_1(\phi,X)^3 ( \alpha_1(\phi,X)+ \alpha_2(\phi,X)X)\:g,
\end{equation}
where $\bar{g} = \det (\bar{g}_{\mu\nu})$ and $g = \det (g_{\mu\nu})$. Let us notice that the assumptions about the functions made so far guarantee that the determinant will not vanish.

There are also some purely physical conditions that the disformal transformation must satisfy. First of all, the transformation must preserve the Lorentzian signature of the metric; this requirement will be fulfilled if \cite{bettoni:2013}:
\begin{equation}
     \alpha_1(\phi,X)+ \alpha_2(\phi,X)X > 0.
\end{equation}
Given that we do not exclude the possibility of $\alpha_2(\phi,X) = 0$, we must have $\alpha_1(\phi,X) > 0$.

Moreover, the transformation must keep the causality properties of the trajectories unchanged. This condition leads us to the following requirements \cite{annala:2021}:
\begin{equation}
    \alpha_1(\phi,X) >0, \quad \alpha_2(\phi, X)\leq 0.
\end{equation}

As it was stated before, we now aim at introducing an additional transformation of the independent, symmetric connection. We thus postulate the following formula:
\begin{equation}\label{eq:transf_kon}
    \bar{\Gamma}^\alpha_{\mu\nu} = \Gamma^\alpha_{\mu\nu} + 2\gamma_1(\phi,X)\delta^\alpha_{(\mu}\phi_{\nu )} + \gamma_2(\phi,X)g_{\mu\nu}\phi^\nu + \gamma_3(\phi,X)\phi_\nu\phi_\nu\phi^\alpha,
\end{equation}
controlled by three arbitrary functions $\gamma_i$ of the scalar field and the kinetic term $X$. This particular form of the transformation formula will become obvious when the action functional is introduced. 

The inverse transformation, accompanied by an inverse disformal transformation of the metric tensor presented here for full generality, reads as:
\begin{equation}
    \Gamma^\alpha_{\mu\nu} = \bar{\Gamma}^\alpha_{\mu\nu}+2\bar{\gamma}_1(\phi,\bar{X})\delta^\alpha_{(\mu}\phi_{\nu )} + \bar{\gamma}_2(\phi,\bar{X})\bar{g}_{\mu\nu}\bar{\phi}^\nu + \bar{\gamma}_3(\phi,\bar{X})\phi_\nu\phi_\nu\bar{\phi}^\alpha,
\end{equation}
with:
\begin{align}
    &\bar{\gamma}_1(\phi, \bar{X})  = -\gamma_1(\phi, X(\bar{X})),\\
    &\bar{\gamma}_2(\phi, \bar{X})  = -\frac{\gamma_2(\phi, X(\bar{X}))}{1-\alpha_2(\phi, X(\bar{X}))X(\bar{X})}, \\
    &\bar{\gamma}_3(\phi, \bar{X})= - \frac{\alpha_1(\phi, X(\bar{X}))\gamma_3(\phi, X(\bar{X})) - \alpha_2(\phi, X(\bar{X}))\gamma_2(\phi, X(\bar{X}))}{1-\alpha_2(\phi, X(\bar{X}))X(\bar{X})}.
\end{align}
Let us notice that if only the transformation of the connection is carried out, i.e., when $\alpha_1 = 1, \alpha_2 = 0$, these relations are reduced to $\bar{\gamma}_i = -\gamma_i$, $i=1,2,3$.

The proposed transformation, in its generality, changes the geodesics and alters the light cones. However, if $\gamma_3 = 0$, then the light cones are preserved; and if additionally $\gamma_2 =0$, the geodesics of the massive particles also remain the same.

\section{Action functional and transformation formulae}\label{sec:3}

A minimal action functional that extends the Palatini KGB parametrization of Horndeski's gravity, being at the same time form-invariant under the transformations \eqref{eq:transf1} and \eqref{eq:transf_kon}, is given as:
\begin{equation}\label{eq:action}
    \begin{split}
        S(g, \Gamma,\phi;\chi) = &\frac{1}{2\kappa^2}\int d^4 x\sqrt{-g}\Big(A_1(\phi, X)g^{\mu\nu}R_{\mu\nu}(\Gamma) + A_2(\phi,X)\phi^\mu\phi^\nu R_{\mu\nu}(\Gamma) + B(\phi, X)\Box\phi + C(\phi,X)\phi^\mu\phi^\nu\nabla_\mu\phi_\nu \\
        &+ E_1(\phi, X)Q_\alpha^{\:\alpha\beta}\phi_\beta + E_2(\phi, X)g_{\mu\nu}Q_\alpha^{\:\mu\nu}\phi^\alpha + E_3(\phi, X)Q_\alpha^{\:\mu\nu}\phi_\mu\phi_\nu\phi^\alpha + F(\phi, X) \Big) \\
        &+ S_\text{matter}(\Sigma_1(\phi,X)g_{\mu\nu} + \Sigma_2(\phi, X)\phi_\mu\phi_\nu;\chi).
    \end{split}
\end{equation}
Here, $\kappa^2 = 8\pi G/c^4$, $\chi$ represent collectively matter fields, $\Box = g^{\mu\nu}\nabla_\mu\nabla_\nu$, the covariant derivative is defined w.r.t the independent connection $\Gamma^\alpha_{\mu\nu}$, $Q_{\alpha}^{\:\mu\nu}$ is the non-metricity tensor defined as $Q_{\alpha}^{\:\mu\nu} = \nabla_\alpha g^{\mu\nu}$. The Ricci tensor is constructed from the independent connection only and thus is unaffected by the transformation of the metric tensor.

Some justification for this particular form of the action is in order. First of all, the inclusion of the terms involving non-metricity are a consequence of the non-uniqueness of the definition of the d'Alembert operator. Since in the metric version of the theory the non-metricity vanishes, there is no ambiguity; however, in the Palatini case, one can come up with different realizations of the operator, of which the most general is:
\begin{equation}
    \Box ^{(P)}:= a_1g^{\mu\nu}\nabla_\mu\nabla_\nu + a_2\nabla_\mu\left(g^{\mu\nu}\nabla_\nu\right) + a_3\nabla_\mu\nabla_\nu \left(g^{\mu\nu}\cdot\right),
\end{equation}
with $a_1 + a_2 +a_3 = 1$, so that the operator reduces to the standard d'Alembert operator if the theory is purely metric. If we use the Leibnitz rule, we will be able to write the above definition as:
\begin{equation}
    \Box ^{(P)}:= g^{\mu\nu}\nabla_\mu\nabla_\nu + (a_2 + 2a_3)Q^\nu\nabla_\nu + a_3 (\nabla_\nu Q^\nu)(\cdot),
\end{equation}
with $Q^\nu = Q_\mu^{\:\mu\nu}$. We can then promote the parameters $a_i$ to functions of the scalar field and the kinetic term, and set $a_3 = 0$ for simplicity; this justifies the addition of the term $E_1(\phi, X)Q_\alpha^{\:\alpha\beta}\phi_\beta$. After adding this particular term, the requirement of the form-invariance of the action under the disformal transformation of the metric tensor forces us to add the remaining terms involving the non-metricity tensor. It also happens that the action \eqref{eq:action} allows for the connection transformation given by \eqref{eq:transf_kon}.

Let us notice that there are additional functions of the scalar field entering the matter sector, leading potentially to breaking of the Weak Equivalence Principle and appearance of the fifth force. The functions were included in order to accommodate the possibility of such an anomalous coupling in a general setup. Moreover, by default, one allows for a kinetic coupling between the matter fields and the scalar field. 

Overall, the action contains 10 functions of the scalar field and the kinetic term. One has to notice, however, that by integrating by parts one can effectively get rid of the $B(\phi, X)$ function and express $\sqrt{-g} B(\phi, X)\Box\phi$ as a combination of terms involving $\phi^\mu\phi^\nu\nabla_\mu\nabla_\nu\phi$, $Q_{\alpha}^{\:\alpha\beta}\phi_\beta$, $g_{\mu\nu}Q_{\alpha}^{\:\mu\nu}\phi^\alpha$, and a function of the field and the kinetic terms, leading to a modification of the terms proportional to $C(\phi, X)$, $E_1(\phi, X)$, $E_2(\phi, X)$, and the function $F(\phi, X)$; therefore, only nine of them are needed to define the theory. Nevertheless, we decide to keep it in the action, as its presence will simplify the transformation formulae introduced below. 

In analogy to the standard ST theories in the Wagoner parametrization, we will refer to a particular choice of the functions as a choice of the disformal frame. Frames can be changed by means of the transformations of the metric and the connection. For simplicity, we will consider the transformation formulae separately for the connection and the metric tensor. If we apply the disformal transformations \eqref{eq:transf_met_inv} to the latter, we get:
\begin{align}\label{eq:transf_for_met}
    & \bar{A}^{(m)}_1(\phi, \bar{X}) = f(\phi, \bar{X})\bar{\beta}_1(\phi, \bar{X}) A_1(\phi, X(\bar{X})), \\
    & \bar{A}^{(m)}_2(\phi, \bar{X}) = f(\phi, \bar{X})\Big(\bar{\beta}_2(\phi, \bar{X})A_1(\phi, X(\bar{X})) + \big(\bar{\beta}_1(\phi, \bar{X}) + \bar{\beta}_2(\phi, \bar{X})\bar{X}\big)^2 A_2(\phi, X(\bar{X}))\Big), \\
    & \bar{B}^{(m)}(\phi, \bar{X}) = f(\phi, \bar{X})\Big(\bar{\beta}_1(\phi, \bar{X}) B(\phi, X(\bar{X})) +\bar{\beta}_2(\phi, \bar{X})\bar{X} E_1(\phi, X(\bar{X})) \Big), \\
    & \bar{C}^{(m)}(\phi, \bar{X}) =  f(\phi, \bar{X})\Big(\big(\bar{\beta}_1(\phi, \bar{X}) + \bar{\beta}_2(\phi, \bar{X})\bar{X}\big)^2 C(\phi, X(\bar{X})) + \bar{\beta}_2(\phi, \bar{X}) B(\phi, X(\bar{X})) \nonumber \\
    & \quad\quad\quad\quad + E_1(\phi, X(\bar{X}))\big(2\bar{\beta}_{1,\bar{X}}(\phi, \bar{X}) + \bar{\beta}_{2}(\phi, \bar{X}) + 2\bar{\beta}_{2,\bar{X}}(\phi, \bar{X})\bar{X}\big) \nonumber \\
   & \quad\quad\quad\quad +2\big(\bar{\beta}_1(\phi, \bar{X}) + \bar{\beta}_2(\phi, \bar{X})\bar{X}\big)E_2(\phi, X(\bar{X}))\big( 4 \bar{\alpha}_1(\phi,\bar{X})\bar{\beta}_{1,\bar{X}}(\phi, \bar{X}) + \bar{\alpha}_2(\phi,\bar{X})\bar{\beta}_{1,\bar{X}}(\phi, \bar{X})\bar{X} \nonumber\\
    & \quad\quad\quad\quad + \bar{\alpha}_1(\phi,\bar{X})\bar{\beta}_{2,\bar{X}}(\phi, \bar{X})\bar{X} + \bar{\alpha}_2(\phi,\bar{X})\bar{\beta}_{2,\bar{X}}(\phi, \bar{X})\bar{X}^2 + \bar{\alpha}_1(\phi,\bar{X})\bar{\beta}_{2}(\phi, \bar{X})+\bar{\alpha}_2(\phi,\bar{X})\bar{\beta}_{2}(\phi, \bar{X})\bar{X}\big)\nonumber \\
   & \quad\quad\quad\quad+ 2\big(\bar{\beta}_1(\phi, \bar{X}) + \bar{\beta}_2(\phi, \bar{X})\bar{X}\big)E_3(\phi, X(\bar{X}))\big(\bar{\beta}_{1,\bar{X}}(\phi, \bar{X}) \bar{X}+\bar{\beta}_{2,\bar{X}}(\phi, \bar{X}) \bar{X}^2 +\bar{\beta}_{2}(\phi, \bar{X}) \bar{X}\big)\Big), \\
   & \bar{E}^{(m)}_1(\phi, \bar{X}) = f(\phi, \bar{X})\Big(\big(\bar{\beta}_1(\phi, \bar{X}) + \bar{\beta}_2(\phi, \bar{X})\bar{X}\big)E_1(\phi, X(\bar{X}))\Big), \\
   &\bar{E}^{(m)}_2(\phi, \bar{X}) = f(\phi, \bar{X})\Big(\big(\bar{\beta}_1(\phi, \bar{X}) + \bar{\beta}_2(\phi, \bar{X})\bar{X}\big)E_2(\phi, X(\bar{X}))\Big), \\
   & \bar{E}^{(m)}_3(\phi, \bar{X}) = f(\phi, \bar{X})\Big(E_1(\phi, X(\bar{X}))\big(\bar{\beta}_{1,\bar{X}}(\phi, \bar{X}) + \bar{\beta}_{2}(\phi, \bar{X}) + \bar{\beta}_{2,\bar{X}}(\phi, \bar{X})\bar{X}\big) \nonumber \\
   & \quad\quad\quad\quad +\big(\bar{\beta}_1(\phi, \bar{X}) + \bar{\beta}_2(\phi, \bar{X})\bar{X}\big)E_2(\phi, X(\bar{X}))\big( 4 \bar{\alpha}_1(\phi,\bar{X})\bar{\beta}_{1,\bar{X}}(\phi, \bar{X}) + \bar{\alpha}_2(\phi,\bar{X})\bar{\beta}_{1,\bar{X}}(\phi, \bar{X}) \bar{X}+ \bar{\alpha}_2(\phi,\bar{X})\bar{\beta}_{1}(\phi, \bar{X})\nonumber\\
    & \quad\quad\quad\quad + \bar{\alpha}_1(\phi,\bar{X})\bar{\beta}_{2,\bar{X}}(\phi, \bar{X})\bar{X} + \bar{\alpha}_2(\phi,\bar{X})\bar{\beta}_{2,\bar{X}}(\phi, \bar{X})\bar{X}^2 +2 \bar{\alpha}_1(\phi,\bar{X})\bar{\beta}_{2}(\phi, \bar{X})+2\bar{\alpha}_2(\phi,\bar{X})\bar{\beta}_{2}(\phi, \bar{X})\bar{X}\big) \nonumber \\
    & \quad\quad\quad\quad +\big(\bar{\beta}_1(\phi, \bar{X}) + \bar{\beta}_2(\phi, \bar{X})\bar{X}\big)E_3(\phi, X(\bar{X}))\big(\bar{\beta}_{1,\bar{X}}(\phi, \bar{X}) \bar{X}+\bar{\beta}_{2,\bar{X}}(\phi, \bar{X}) \bar{X}^2 +2\bar{\beta}_{2}(\phi, \bar{X}) \bar{X} + \bar{\beta}_{1}(\phi, \bar{X})\big),\\
    & \bar{F}^{(m)}(\phi, \bar{X}) = f(\phi, \bar{X})\Big(F(\phi, X(\bar{X})) +E_3(\phi, X(\bar{X}))\big(\bar{\beta}_1(\phi, \bar{X}) + \bar{\beta}_2(\phi, \bar{X})\bar{X}\big)\big(\bar{\beta}_{1,\phi}(\phi, \bar{X}) + \bar{\beta}_{2,\phi}(\phi, \bar{X})\bar{X}\big)\bar{X}^2 \nonumber \\
    & \quad\quad\quad\quad +   E_2(\phi, X(\bar{X}))\big(\bar{\beta}_1(\phi, \bar{X}) + \bar{\beta}_2(\phi, \bar{X})\bar{X}\big)\big(4 \bar{\alpha}_1(\phi,\bar{X})\bar{\beta}_{1,\phi}(\phi, \bar{X})\bar{X} + \bar{\alpha}_2(\phi,\bar{X})\bar{\beta}_{1,\phi}(\phi, \bar{X})\bar{X}^2 \nonumber\\
    & \quad\quad\quad\quad + \bar{\alpha}_1(\phi,\bar{X})\bar{\beta}_{2,\phi}(\phi, \bar{X})\bar{X}^2 + \bar{\alpha}_2(\phi,\bar{X})\bar{\beta}_{2,\phi}(\phi, \bar{X})\bar{X}^3) + E_1(\phi, X(\bar{X}))\big(\bar{\beta}_{1,\phi}(\phi, \bar{X}) + \bar{\beta}_{2,\phi}(\phi, \bar{X})\bar{X}\big)\bar{X}\Big),\\
    & \bar{\Sigma}_1^{(m)}(\phi, \bar{X}) = \bar{\alpha}_1(\phi, \bar{X})\Sigma_1(\phi, X(\bar{X})), \\
    & \bar{\Sigma}_2^{(m)}(\phi, \bar{X}) = \Sigma_2(\phi, X(\bar{X}))+\bar{\alpha}_2(\phi, \bar{X})\Sigma_1(\phi, X(\bar{X})),
\end{align}
where $f(\phi, \bar{X})=\bar{\alpha}^{3/2}_1(\phi,\bar{X})\sqrt{\bar{\alpha}_1(\phi,\bar{X})+\bar{\alpha}_2(\phi,\bar{X})\bar{X}}$, $\bar{\beta}_{i, \phi}$ denotes derivative w.r.t. the scalar field, $\bar{\beta}_{i, \bar{X}}$ stands for the derivative w.r.t. the new kinetic term $\bar{X}$, and superscript $^{(m)}$ means that the transformation has been carried out only for the metric tensor. 

Analogously, under the change of the connection given by Eq. \eqref{eq:transf_kon}, we obtain the following formulae:
\begin{align}\label{eq:transf_form_conn}
    & \bar{A}_1^{(c)}(\phi,X) = A_1(\phi,X), \\
    & \bar{A}_2^{(c)}(\phi,X) = A_2(\phi,X), \\
    & \bar{B}^{(c)}(\phi,X) = B(\phi,X)-3A_1(\phi,X)\big(\bar{\gamma}_1(\phi,X)-\bar{\gamma}_2(\phi,X)\big) + A_2(\phi,X)\big(\bar{\gamma}_3(\phi,X) X^2+\bar{\gamma}_2(\phi,X)X\big), \\
    & \bar{C}^{(c)}(\phi,X) = C(\phi,X) -6A_1(\phi,X)\big(\bar{\gamma}_{1,X}(\phi,X) -\bar{\gamma}_{2,X}(\phi,X) \big)\nonumber \\
    &\quad\quad\quad\quad-A_2(\phi,X)\big(3(\bar{\gamma}_1(\phi,X)+2\bar{\gamma}_{1,X} (\phi,X)) + \bar{\gamma}_2(\phi,X)+\bar{\gamma}_3(\phi,X)X\big), \\
    & \bar{E}_1^{(c)}(\phi,X) = E_1(\phi, X) + A_1(\phi, X)\big(4\bar{\gamma}_2(\phi, X)+\bar{\gamma}_3(\phi, X)X\big) + A_2(\phi,X)\big(\bar{\gamma}_2(\phi, X)X+\bar{\gamma}_3(\phi, X)X^2\big), \\
    &  \bar{E}_2^{(c)}(\phi,X) = E_2(\phi, X)- A_1(\phi,X)\bar{\gamma}_2(\phi, X), \\
    & \bar{E}_3^{(c)}(\phi,X) = E_3(\phi, X) - A_1(\phi, X)\big(3(\bar{\gamma}_1(\phi,X)-\bar{\gamma}_2(\phi,X)) + \bar{\gamma}_3(\phi,X)\big) \nonumber \\
    & \quad\quad\quad\quad-A_2(\phi,X)\big(3\bar{\gamma}_{1,X} (\phi,X)X + \bar{\gamma}_2(\phi,X)+\bar{\gamma}_3(\phi,X)X\big), \\
    & \bar{F}^{(c)}(\phi,X) = F(\phi, X) - B(\phi,X)\big(2\bar{\gamma}_1(\phi,X)X + 4\bar{\gamma}_2(\phi,X)X+\bar{\gamma}_3(\phi,X)X^2\big) \nonumber \\
    & \quad\quad\quad\quad - C(\phi,X)\big(2\bar{\gamma}_1(\phi,X)X + \bar{\gamma}_2(\phi,X)X+\bar{\gamma}_3(\phi,X)X^3\big)\nonumber \\
    & \quad\quad\quad\quad + E_1(\phi,X)X \big(7\bar{\gamma}_1(\phi,X) + 5\bar{\gamma}_2(\phi,X) + 2\bar{\gamma}_3(\phi,X)X\big)\nonumber \\
    & \quad\quad\quad\quad + E_2(\phi,X)X \big(10\bar{\gamma}_1(\phi,X) + 2\bar{\gamma}_2(\phi,X) + 2\bar{\gamma}_3(\phi,X)X\big)\nonumber \\
    & \quad\quad\quad\quad + E_3(\phi,X)X^2 \big(4\bar{\gamma}_1(\phi,X) + 2\bar{\gamma}_2(\phi,X) + 2\bar{\gamma}_3(\phi,X)X\big)\nonumber \\
    & \quad\quad\quad\quad + A_1(\phi,X)X\big(-3\big(\bar{\gamma}_{1,\phi}(\phi,X) -\bar{\gamma}_{2,\phi}(\phi,X) - \bar{\gamma}^2_{1}(\phi,X) -\bar{\gamma}^2_{2}(\phi,X)-4\bar{\gamma}_{1}(\phi,X)\bar{\gamma}_{2}(\phi,X)\big) \nonumber \\
    & \quad\quad\quad\quad +3X(\bar{\gamma}_{1}(\phi,X)+\bar{\gamma}_{2}(\phi,X))\bar{\gamma}_{3}(\phi,X)\big)\nonumber \\
    &  \quad\quad\quad\quad + 3A_2(\phi,X)X^2\big(\bar{\gamma}^2_{1}(\phi,X)-\bar{\gamma}_{1,\phi}(\phi,X) +\bar{\gamma}_{1}(\phi,X)\bar{\gamma}_{2}(\phi,X) + \bar{\gamma}_{1}(\phi,X)\bar{\gamma}_{3}(\phi,X)X\big), \\
    & \bar{\Sigma}_1^{(c)}(\phi, X) = \Sigma_1(\phi, X), \\
    & \bar{\Sigma}_2^{(c)}(\phi, X) = \Sigma_2(\phi, X).
\end{align}
Here, $^{(c)}$ denotes transformation of the independent connection only. For this reason, on both sides of the equality signs we have the same kinetic term $X$. 

One has to notice that, in principle, a reparametrization of the scalar field (i.e., a diffeomorphism $\bar{\phi} = f(\phi)$) should also be allowed. This would not, however, influence the main results of the paper, and at the same time would contribute to a considerably increased level of complexity of the transformation formulae.

It is now necessary to give an explicit formula for compositions of two consecutive transformations of the metric and the connection. Let us assume that we start from a frame with the metric $\bar{\bar{g}}_{\mu\nu}$, the connection $\bar{\bar{\Gamma}}^\alpha_{\mu\nu}$, the kinetic term $\bar{\bar{X}}$, and the 10 functions of the scalar field. Next, we transform the metric and the connection using the five functions $\{\bar{\bar{\alpha}}_i(\phi, \bar{X}), \bar{\bar{\gamma}}_j(\phi, \bar{X})\}$, and arrive at a new disformal frame with the variables $(\bar{g}_{\mu\nu}, \bar{\Gamma}^\alpha_{\mu\nu})$, and the kinetic term $\bar{X}$. We carry out one more transformation controlled by $\{{\bar{\alpha}}_i(\phi, X), {\bar{\gamma}}_j(\phi, X)\}$ in order to end up in a frame with $({g}_{\mu\nu}, {\Gamma}^\alpha_{\mu\nu})$ and the kinetic term $X$. This situation is presented in diagram \eqref{diag:comm}.
\begin{equation}\label{diag:comm}
    \begin{tikzcd}[row sep=huge, column sep = huge]
    (\bar{\bar{g}}, \bar{\bar{\Gamma}}) \arrow{r}{\{\bar{\bar{\alpha}}_i,\bar{\bar{\gamma}}_j\}} \arrow[swap]{dr}{\{{{\alpha}}_i,{{\gamma}}_j\}} & (\bar{g}, \bar{\Gamma}) \arrow{d}{\{{\bar{\alpha}}_i,{\bar{\gamma}}_j\}} \\
     & (g, \Gamma)
  \end{tikzcd}
\end{equation}
 We now ask a question: what should the five transforming functions be if we were to transform directly from the frame with $(\bar{\bar{g}}_{\mu\nu},\bar{\bar{\Gamma}}^\alpha_{\mu\nu})$ to $({g}_{\mu\nu}, {\Gamma}^\alpha_{\mu\nu})$, using $\{{{\alpha}}_i(\phi, X), {{\gamma}}_j(\phi, X)\}$? The functions defining the disformal transformation transform as follows:
 \begin{align}\label{eq:comp_met}
    & \alpha_1(\phi, X) = \bar{\bar{\alpha}}_1(\phi, {\bar{X}}(X)){\bar{\alpha}}_1(\phi,X), \\
    & \alpha_2(\phi, X) =\bar{\bar{\alpha}}_1(\phi, {\bar{X}}(X)){\bar{\alpha}}_2(\phi, X)+\bar{\bar{\alpha}}_2(\phi, {\bar{X}}(X)),
 \end{align}
with $\bar{X}= X/(\bar{\alpha}_1(\phi,X) + \bar{\alpha}_2(\phi,X)X)$ and $\bar{\bar{X}}= \bar{X}/(\bar{\bar{\alpha}}_1(\phi,\bar{X})+ \bar{\bar{\alpha}}_2(\phi,\bar{X})\bar{X})$. The functions defining the transformation of the connection are given by:
\begin{align}
    &\gamma_1(\phi, X) = \bar{\gamma}_1(\phi, X) + \bar{\bar{\gamma}}_1(\phi, {\bar{X}}(X)),\\
    & \gamma_2(\phi, X) = \bar{\gamma}_2(\phi, X) + \bar{\bar{\gamma}}_2(\phi, {\bar{X}}(X)) +\bar{\bar{\gamma}}_2(\phi, {\bar{X}}(X))\bar{\alpha}_1(\phi,X)\bar{\beta}_2(\phi,X) X, \\
    & \gamma_3(\phi, X) = \bar{\gamma}_3(\phi, X) + \bar{\bar{\gamma}}_2(\phi, \bar{X}(X))\bar{\alpha}_2(\phi,X)\big(\bar{\beta}_1(\phi,X) + \bar{\beta}_2(\phi,X) X\big) + \bar{\bar{\gamma}}_3(\phi, \bar{X}(X))\big(\bar{\beta}_1(\phi,X) + \bar{\beta}_2(\phi,X) X\big).
\end{align}
Let us denote the action of the transformations \eqref{eq:transf1} and \eqref{eq:transf_kon} on the metric tensor and the connection by $(\{\alpha_1, \alpha_2, \gamma_1, \gamma_2,\gamma_3\}\equiv)\{\alpha, \gamma \}\triangleright (g, \Gamma)$, and composition of two such transformations by $\{\bar{\bar{\alpha}}, \bar{\bar{\gamma}}\}\circ \{{\bar{\alpha}}, {\bar{\gamma}}\} \equiv \{\alpha, \gamma\}$. The neutral element will be denoted by $\{0, 0\}$ and is given by $\{\alpha_1=1, \alpha_2=0, \gamma_1=0,\gamma_2=0, \gamma_3=0\}$. Then, by inspecting the formulas given above, we can make the following observation:
\begin{equation}
   ( \{\alpha, 0\}\circ \{0, \gamma\})\triangleright (g, \Gamma)\neq (\{0, \gamma\} \circ \{\alpha, 0\} )\triangleright (g, \Gamma).
\end{equation}
In other words, the order of the transformations does matter, in the sense that the application of two consecutive transformations given by the same functional dependence of $\phi$ and $X$ will yield two different results, depending on their order. It must be noted, though, that there exist commutative subgroups (apart from trivial cases like $\{\alpha_1, 0,0,0,0\}$), such as those constructed from the elements of the form $\{\alpha_1, 0, \gamma_1, \gamma_2, 0\}$. 

Taking into account everything that has been said about the transformations, we can now describe the procedure of coming up the full transformations of the scalar field functions entering the action functional, i.e., of relating these functions expressed in a frame with $(g_{\mu\nu}, \Gamma^\alpha_{\mu\nu})$ to the ones from a frame defined by $(\bar{g}_{\mu\nu}, \bar{\Gamma}^\alpha_{\mu\nu})$, with both the metric and the connection changed. We can do that either by transforming first the metric and then the connection, or doing the opposite. This situation is schematically represented in diagram \eqref{diag:full_transf}:

\begin{equation}\label{diag:full_transf}
    \begin{tikzcd}
(\bar{g}, \bar{\Gamma}) \arrow{r}{\{\alpha, 0\}} \arrow[swap]{d}{\{0, \bar{\gamma}'\}} &(g, \bar{\Gamma}) \arrow{d}{\{0, \bar{\gamma}\}} \\%
(\bar{g}, \Gamma) \arrow{r}{\{\alpha, 0\}}& (g, \Gamma)
\end{tikzcd}
\end{equation}

which will commute if:
\begin{align}
    & \bar{\gamma}_1(\phi, X) = \bar{\gamma}'_1(\phi, \bar{X}(X)), \\
    & \bar{\gamma}_2(\phi, X) = \bar{\gamma}'_2(\phi, \bar{X}(X))\big(1 + \alpha_1(\phi, X)\beta_2(\phi, X) X\big), \\
    & \bar{\gamma}_3(\phi, X) = \bar{\gamma}'_2(\phi, \bar{X}(X))\alpha_2(\phi,X)\big(\beta_1(\phi, X) +\beta_2(\phi, X) X\big) + \bar{\gamma}'_3(\phi, \bar{X}(X))\big(\beta_1(\phi, X) +\beta_2(\phi, X) X\big).
\end{align}
The prime denotes only alternative transformation functions, not a derivative. We will not present here the result of a simultaneous transformation of the metric and the connection (i.e., of $\{\alpha, \gamma\}\triangleright (g, \Gamma)$), as it would not be particularly illuminating.

\subsection{Related MA theories}

Here, we will establish direct links between the proposed action functional and various MA ST theories of gravity analyzed so far in the literature:
\begin{itemize}
    \item Scalar tensor theory of gravity in the Palatini approach analyzed in Ref. \cite{kozak:2019}:
    \begin{equation}
    \begin{split}
       S &  = \frac{1}{2\kappa^2}\int d^4x \sqrt{-g}\Big(\mathcal{A}(\phi) g^{\mu\nu}R_{\mu\nu}(\Gamma) -\mathcal{B}(\phi)X - \mathcal{V}(\phi) + \mathcal{C}_1(\phi)g_{\mu\nu}\phi^\alpha Q_\alpha^{\:\mu\nu} -\mathcal{C}_2(\phi)Q^\alpha \phi_\alpha\Big) \\
       & \quad\quad\quad + S_m\left(e^{2\alpha(\phi)}g_{\mu\nu};\chi\right),
    \end{split}
    \end{equation}
    with the following identifications: $A_1(\phi,X) = \mathcal{A}(\phi)$, $A_2(\phi, X) = 0$, $B(\phi,X) = 0$, $C(\phi,X) = 0$, $E_1(\phi,X) = -\mathcal{C}_2(\phi)$, $E_2(\phi,X) =\mathcal{C}_1(\phi)$, $E_3(\phi,X) = 0$, $F(\phi,X) = -\mathcal{B}(\phi)X - \mathcal{V}(\phi)$. Alternatively, one can notice that it is possible to consider non-zero $B(\phi)$ function (with no dependence on $X$) in the original action \eqref{eq:action}, and integrate by parts the term $\sqrt{-g}B(\phi)\Box\phi$, leading to different $E_1(\phi, X)$ and $E_1(\phi, X)$ in the general theory (and influencing the $F(\phi, X)$ function as well). 

    Similar actions, but without the non-metricity, have been analyzed often in the literature, particularly in the context of cosmic inflation; see Ref. \cite{racioppi:2017, racioppi:2018, jarv:2018, gialamas:2023, bauer:2008,rasanen:2017,antoniadis:2018,enckell:2019}. It must be also noted that these theories contain Palatini $f(R)$ gravity, which can be represented as ST theories by means of the Legendre transformation \cite{olmo:2011}. 
    \item Theories with kinetic coupling to the curvature, e.g.:
    \begin{equation}
        \begin{split}
            S & = \frac{1}{2\kappa^2}\int d^4x\sqrt{-g}\Big(\frac{1}{2}F_1(\phi, X) g^{\mu\nu}R_{\mu\nu}(\Gamma) + F_3(\phi, X)\phi^\mu\phi^\nu R_{\mu\nu}(\Gamma)-\frac{1}{2}F_2(\phi,X)X+\frac{1}{4}F_4(\phi,X)X^2\\
            & \quad\quad\quad-F_5(\phi,X)V(\phi)\Big),
        \end{split}
    \end{equation}
    with the matter part neglected, analyzed in Ref. \cite{gialamas:2020}. Here, the identifications are: $A_1(\phi,X) = \frac{1}{2}F_1(\phi, X) $, $A_2(\phi, X) = F_3(\phi, X)$, $B(\phi,X) = 0$, $C(\phi,X) = 0$, $E_1(\phi,X) = 0$, $E_2(\phi,X) =0$, $E-3(\phi,X) = 0$, $F(\phi,X) = -\frac{1}{2}F_2(\phi,X)X+\frac{1}{4}F_4(\phi,X)X^2-F_5(\phi,X)V(\phi)$. Similar actions have been analyzed in Ref. \cite{galtsov:2019, galtsov:2020, luo:2014}. In the metric approach, such actions have been analyzed somewhat more often, e.g. in Ref. \cite{amendola:1993, capozziello:1999,yi:2017,ema:2015,myung:2016,tumurtushaa:2019,sushkov:2009,zhang:2025,sushkov:2023,fatykhov:2025,oikonomou:2024}
    \item Palatini-Horndeski theories, considered in Ref. \cite{helpin:2020a}:
    \begin{equation}
        \begin{split}
            S = & \frac{1}{2\kappa^2}\int d^4x\sqrt{-g}\Big(G_4(\phi)g^{\mu\nu}R_{\mu\nu}(\Gamma)+G_3(\phi, X)\Box^{(m)}\phi + C_1(\phi, X)g_{\mu\nu} Q_\alpha^{\:\mu\nu}\phi^\alpha - C_2(\phi, X)Q_{\alpha}^{\:\alpha\beta}\phi_\beta \\
            & \quad\quad- C_3(\phi, X)Q_\alpha^{\:\mu\nu}\phi^\alpha \phi_\mu\phi_\nu + C_4(\phi, X) \phi_\alpha T^\alpha+ G_2(\phi, X)\Big),
        \end{split}
    \end{equation}
    where $\Box^{(m)}$ is the d'Alembert operator defined w.r.t the metric tensor, $T^\alpha = g^{\alpha\beta} T^{\mu}_{\:\beta\mu}$, $T^\mu_{\:\beta\mu}=2\Gamma^\mu_{\:[\beta\mu]}$ is the torsion tensor; the matter part was omitted. Because of the form of the action and the existence of the torsion, the direct comparison with the action \eqref{eq:action} is not possible; one would have to assume that the connection is symmetric. Moreover, in order to express the metric d'Alembert operator as the metric-affine, one could use the fact that the independent connection could be related to the Levi-Civita connection of the metric tensor $\mathring{\Gamma}^\alpha_{\:\mu\nu}$ through $\Gamma^\alpha_{\:\mu\nu} = \mathring{\Gamma}^\alpha_{\:\mu\nu} + C^\alpha_{\:\mu\nu}$, where $C^\alpha_{\:\mu\nu}$ is some tensor.
    \item Small-$\tilde{\kappa}$ \footnote{In the literature, the EiBI theory parameter is denoted by $\kappa$, without the bar, but in this paper we use this symbol in a way it was defined earlier on.} limit of Eddington-inspired Born-Infeld (EiBI) gravity with a scalar field included \cite{pani:2012, banados:2010}. The action in the Jordan frame is:
    \begin{equation}
        S^{(J)} = \frac{2}{\tilde{\kappa}}\int d^4 x\big(\sqrt{|g_{\mu\nu} + \tilde{\kappa} R_{\mu\nu}(\Gamma)|} - \lambda \sqrt{-g}\big) + S_m(g_{\mu\nu}; \chi),
    \end{equation}
    where $\tilde{\kappa}$ is the theory parameter, and $\lambda$ is related to the cosmological constant $\Lambda$, $\Lambda=(\lambda- 1)/\tilde{\kappa}$. One can transform the action to the Einstein frame by introducing the metric $\gamma_{\mu\nu} = g_{\mu\nu} + \tilde{\kappa} R_{\mu\nu}(\Gamma)$:
    \begin{equation}
        \begin{split}
            S^{(E)} = & \frac{-2\lambda}{\tilde{\kappa}}\int d^4 x\big(\sqrt{|\gamma_{\mu\nu} - \tilde{\kappa} R_{\mu\nu}(\Gamma)|} - \frac{1}{\lambda}\sqrt{-\gamma}\big) +S_m(\gamma_{\mu\nu} -\tilde{\kappa} R_{\mu\nu}(\Gamma); \chi).
        \end{split}
    \end{equation}
    with $\gamma$ being the determinant of the metric $\gamma_{\mu\nu}$. Taking the Lagrangian for the scalar field:
    \begin{equation}
        \mathcal{L}_m(g_{\mu\nu}; \phi) = g^{\mu\nu}\phi_\mu\phi_\nu,
    \end{equation}
    and limit of small $\tilde{\kappa}$, one arrives at:
    \begin{equation}
        S^{(E)} = \int d^4 x \sqrt{-\gamma} \Big(\lambda \gamma^{\mu\nu}R_{\mu\nu}(\Gamma) - 2\frac{\lambda - 1}{\tilde{\kappa}} + \gamma^{\mu\nu}\phi_\mu\phi_\nu - \tilde{\kappa} \big(R_{\mu\nu}(\Gamma) + \frac{1}{2}\gamma_{\mu\nu}\gamma^{\alpha\beta}R_{\alpha\beta}(\Gamma)\big) \phi^\mu\phi^\nu+ \mathcal{O}(\tilde{\kappa}^2) \Big).
    \end{equation}
    Obviously, this action corresponds to the original one (c.f. \eqref{eq:action}) with the following identifications: $A_1(\phi, X) = \lambda$. $A_2(\phi, X) = -\bar{\kappa}(1 - X^{(\gamma)})$, where $X^{(\gamma)} = \gamma^{\mu\nu}\phi_\mu\phi_\nu$, $F(\phi, X) = X^{(\gamma)} - 2\frac{\lambda - 1}{\bar{\kappa}}$, and the remaining functions vanish. 
\end{itemize}

\subsection{Invariant metric and connection}

The possibility of introducing a conformal mapping of the metric tensor in the case of ST theories poses the difficulty of identifying the conformal frame which should be considered as physical; the literature on this topic is extensive, examples are given in Ref. \cite{flanagan:2004, faraoni:1999a, faraoni:1999b}. The two most commonly used frames are the Einstein and the Jordan, depending on whether the scalar field is coupled non-minimally to the curvature or to the matter part of the action. Different arguments have been given in favour of either of them; see previous references and Ref. \cite{quiros:2019}. Some authors, however, claim that the issue of choosing the right conformal frame is analogous to trying to find the right coordinate system when describing physical phenomena; obviously, there is no such thing, and for this reason, the description of the laws of nature should be independent of the particular reference frame. Physical quantities should be expressed as invariant combinations of the mathematical objects, and the basic laws of physics should be written in the same way in all coordinates. Similarly, one needs to try to use such quantities in ST theories that are invariant under the conformal transformation to rule out any dependence on the particular frame. One can thus come up with invariant scalar field and invariant metric \cite{jarv:2015}, but also, in the case of the Palatini theory, the invariant connection \cite{kozak:2019}. The invariance of these quantities simply means that if in all frames related by a conformal transformation one decides to choose a particular combination of the metric, the connection, and the scalar field functions, then one will measure the same distances, determine the same geodesics, and so on. Invariant quantities are also useful for establishing mathematical equivalence of different conformal frames; one can, for example, use them in order to find out whether a given ST theory is equivalent to some $f(R)$ gravity, in the metric or in the Palatini approach \cite{kozak:2021}. 

Analogously, one can try to find a metric tensor and a connection that would be invariant under the disformal transformation and the connection mapping defined by Eqs. \eqref{eq:transf1} and \eqref{eq:transf_kon}. Most likely, there is no unique way of doing so, just like in the case of ST theories in the Wagoner parametrization \cite{wagoner:1970}. Of special interest for us, however, will be the following invariant metric:
\begin{equation}\label{eq:inv_metric}
    \hat{g}_{\mu\nu} = \sqrt{A_1(\phi, X)\big(A_1(\phi, X)+A_2(\phi, X)X\big)}g_{\mu\nu} -\frac{\sqrt{A_1(\phi, X)}A_2(\phi, X)}{\sqrt{A_1(\phi, X)+A_2(\phi, X)X}}\phi_\mu\phi_\nu.
\end{equation}
Its invariance can be established by direct calculation, i.e., one can show that, for any metric $\bar{g}_{\mu\nu} = \bar{\alpha}_1(\phi, X)g_{\mu\nu} + \bar{\alpha}_2(\phi, X)\phi_\mu\phi_\nu$, one has:
\begin{equation}
    \begin{split}
         & \sqrt{A_1(\phi, X)\big(A_1(\phi, X)+A_2(\phi, X)X\big)}g_{\mu\nu} -\frac{\sqrt{A_1(\phi, X)}A_2(\phi, X)}{\sqrt{A_1(\phi, X)+A_2(\phi, X)X}}\phi_\mu\phi_\nu =\\
        & = \sqrt{\bar{A}_1(\phi, \bar{X})\big(\bar{A}_1(\phi, \bar{X})+\bar{A}_2(\phi, \bar{X})\bar{X}\big)}\bar{g}_{\mu\nu} -\frac{\sqrt{\bar{A}_1(\phi, \bar{X})}\bar{A}_2(\phi, \bar{X})}{\sqrt{\bar{A}_1(\phi, \bar{X})+\bar{A}_2(\phi, \bar{X})\bar{X}}}\phi_\mu\phi_\nu,
    \end{split}
\end{equation}
where $\bar{A}_i$ denotes the function in the frame where the metric $\bar{g}_{\mu\nu}$ is used. The invariant metric induces an invariant kinetic term:
\begin{equation}
    \hat{X} = \hat{g}^{\mu\nu}\phi_\mu\phi_\nu = \frac{\sqrt{A_1(\phi, X) + A_2(\phi, X)X}}{A_1(\phi, X)^{3/2}}X.
\end{equation}
Furthermore, one can introduce the following invariant connection:
\begin{equation}
    \begin{split}
        \hat{\Gamma}^\alpha_{\mu\nu} = \Gamma^{\alpha}_{\mu\nu} + \mathcal{G}_1(\phi, {X})\big(\delta^\alpha_\nu \phi_\mu + \delta^\alpha_\mu \phi_\nu\big) + \mathcal{G}_2(\phi, {X})\hat{g}_{\mu\nu}\hat{\phi}^\alpha + \mathcal{G}_3(\phi, {X})\phi_\mu\phi_\nu\hat{\phi}^\alpha,
    \end{split}
\end{equation}
where $\hat{\phi}^\alpha = \hat{g}^{\alpha\beta}\phi_\beta$, and the functions $\mathcal{G}_i$ are defined as follows:
{\small
\begin{align}
    & \mathcal{G}_1(\phi, X) = -\frac{3 A_2(\phi, X)X\big(B(\phi, X) - E_1(\phi, X)\big) + A_1(\phi, X)\big(4B(\phi, X) +  C(\phi, X)X -2E_1(\phi, X) + 4E_2(\phi, X) - 2 E_3(\phi, X)X\big)}{12A_1(\phi, X)\big(A_1(\phi, X) + A_2(\phi, X)X\big)}, \\
    & \mathcal{G}_2(\phi, X) = -\frac{-A_2(\phi, X)X\big(B(\phi, X) - E_1(\phi, X)\big) +  A_1(\phi, X)\big(C(\phi, X)X - 2E_1(\phi, X)-4E_2(\phi, X)-2E_3(\phi, X)X\big)}{4A_1(\phi, X)\big(A_1(\phi, X)+A_2(\phi, X)X\big)}, \\
    & \mathcal{G}_3(\phi, X) = \frac{\sqrt{A_1(\phi, X)} \big(-A_2(\phi, X)(B(\phi, X)-E_1(\phi, X)) + A_1(\phi, X)(C(\phi, X)-2E_3(\phi, X))\big)}{2(A_1(\phi, X)+A_2(\phi, X)X)^{3/2}}.
\end{align}}
Each of the functions is invariant under the disformal transformation of the metric tensor; the tensorial constituents are obviously invariant as well, since they are explicitly built from the invariant metric. Moreover, under the change of the connection $\Gamma^\alpha_{\mu\nu} = \bar{\Gamma}^\alpha_{\mu\nu} + 2\bar{\gamma}_1 \delta^\alpha_{(\mu}\phi_{\nu)} + \bar{\gamma}_2 g_{\mu\nu}\phi^\alpha + \bar{\gamma}_3\phi_\mu\phi_\nu\phi^\alpha$, the functions $\mathcal{G}_i$ change as $\mathcal{G}_i = \bar{\mathcal{G}}_i - \bar{\gamma}_i$, so that the terms proportional to $\bar{\gamma}_i$ are canceled out. Notice that the functions depend on the kinetic term as defined in a particular frame, not on the invariant kinetic terms.

\section{Equations of motion and dynamically equivalent metric theory}\label{sec:4}

The field equations for the metric $g_{\mu\nu}$ and the scalar field $\phi$ are extremely complicated in the most general case. One might thus use the freedom of utilizing the five functions $\{\alpha_i, \gamma_j\}$ in order to, for example, annihilate some of the scalar field functions present in the action functional \eqref{eq:action}. Alternatively, one can investigate the field equation for the connection in the hope of integrating it out. One can easily show that the equation resulting from varying the action w.r.t. the independent connection reads as:
\begin{equation}
    \begin{split}
        \nabla_\alpha & \big[\sqrt{-g}\big(A_1(\phi,X)g^{\sigma (\mu} + A_2(\phi,X)\phi^\sigma\phi^{(\mu}\big) \big(\delta^{\nu )}_\sigma\delta^\alpha_\lambda - \delta^{\nu )}_\lambda\delta^\alpha_\sigma\big)\big] = \\
        & = \sqrt{-g}\left((E_1(\phi,X)-B(\phi,X))\phi_\lambda g^{\mu\nu} + (2E_3(\phi,X) - C(\phi,X))\phi_\lambda\phi^\mu\phi^\nu + 2\Big(\frac{1}{2}E_1(\phi,X)+E_2(\phi,X)\Big)\delta^{(\mu}_\lambda\phi^{\nu )}\right),
    \end{split}
\end{equation}
which can be recast in the following form:
\begin{align}\label{eq:conn_fe_1}
     \nabla_\lambda & \big[\sqrt{-g}\big(A_1(\phi,X)g^{ \mu\nu} + A_2(\phi,X)\phi^\mu\phi^{\nu}\big)\big] = \nonumber \\
     & = \sqrt{-g}\Big((E_1(\phi,X)-B(\phi,X))\phi_\lambda g^{\mu\nu} + (2E_3(\phi,X) - C(\phi,X))\phi_\lambda\phi^\mu\phi^\nu \\
     & + 2\Big(\frac{B(\phi,X)}{3}-\frac{2}{3}E_1(\phi,X)-\frac{2}{3}E_2(\phi,X)-\frac{X}{3}(2E_3(\phi,X)-C(\phi,X))\Big)\delta^{(\mu}_\lambda\phi^{\nu )}\Big).
\end{align}
It can be shown that:
\begin{equation}
    \sqrt{-g}\big(A_1(\phi,X)g^{ \mu\nu} + A_2(\phi,X)\phi^\mu\phi^{\nu}\big) = \sqrt{-\hat{g}}\hat{g}^{\mu\nu},
\end{equation}
where $\hat{g}$ is the determinant of the invariant metric \eqref{eq:inv_metric}, and $g^{\mu\nu}$ its inverse. Then, expressing the r.h.s. of Eq. \eqref{eq:conn_fe_1} in terms of this metric and the invariant kinetic term, i.e., by assuming that each function becomes now a function of $\hat{X}$ through its dependence on $X$, we get:
\begin{align}
    \nabla_\lambda \left[\sqrt{-\hat{g}}\hat{g}^{\mu\nu}\right] =\sqrt{-\hat{g}}\Big( k_1(\phi, \hat{X})\delta^{(\mu}_\lambda\hat{\phi}^{\nu )} + k_2(\phi,\hat{X})\hat{g}^{\mu\nu}\phi_\lambda + k_3(\phi, \hat{X})\phi_\lambda\hat{\phi}^\mu\hat{\phi}^\nu\Big),
\end{align}
with:
\begin{align}
    k_1(\phi, \hat{X}) = & \frac{B(\phi ,X(\hat{X}))+ C(\phi ,X(\hat{X}))X(\hat{X})-2 \big(E_1(\phi ,X(\hat{X}))+E_2(\phi ,X(\hat{X}))+
   E_3(\phi ,X(\hat{X}))X(\hat{X})\big)}{3 (A_1(\phi ,X(\hat{X}))+ A_2(\phi ,X(\hat{X}))X(\hat{X}))}, \\
    k_2(\phi, \hat{X}) = & \frac{E_1(\phi ,X(\hat{X}))-B(\phi ,X(\hat{X}))}{A_1(\phi ,X(\hat{X}))}, \\
    k_3(\phi, \hat{X}) = & -2 \mathcal{G}_3(\phi, X(\hat{X})).
\end{align}
Let us notice that we have enough freedom, using the functions $\gamma_i$, to transform the scalar field functions to a frame in which all these three coefficients vanish. In principle, it is enough to find such functions $\gamma_i$ that would result in the following condition for the new frame functions:
\begin{equation}
    \bar{E}_1(\phi, \hat{X}) = \bar{B}(\phi, \hat{X}), \quad \bar{C}(\phi, \hat{X}) = 2\bar{E}_3(\phi, \hat{X}), \quad \bar{E}_2(\phi, \hat{X}) = -\frac{1}{2}\bar{E}_1(\phi, \hat{X}).
\end{equation}
Interestingly, the three functions defining the transformation of the connection turn out to be:
\begin{equation}
    \gamma_1(\phi, \hat{X}) = \mathcal{G}_1(\phi, X(\hat{X})), \quad \gamma_2(\phi, \hat{X}) = \mathcal{G}_2(\phi, X(\hat{X})), \quad \gamma_3(\phi, \hat{X}) = \mathcal{G}_3(\phi, X(\hat{X})),
\end{equation}
i.e., the invariant connection $\hat{\Gamma}$ is the Levi-Civita connection of the invariant metric tensor: $\hat{\Gamma}=\hat{\Gamma}(\hat{g})$, so that, just like in the case of Palatini $f(R)$ gravity, there are no additional degrees of freedom introduced by the connection. In that particular frame, the action functional looks as follows:
\begin{equation}\label{eq:inv_act_ein}
\begin{split}
   S(\hat{g}, \phi;\chi) = & \frac{1}{2\kappa^2}\int d^4 x\sqrt{-\hat{g}}\big(\hat{R}(\hat{g}) + \hat{B}(\phi, \hat{X})\hat{\Box}\phi + \hat{C}(\phi, \hat{X})\hat{\phi}^\mu\hat{\phi}^\nu\hat{\nabla}_\mu\hat{\nabla}_\nu \phi + \hat{F}(\phi, \hat{X})\big) \\
   & + S_m\big(\hat{\Sigma}_1(\phi, \hat{X})\hat{g}_{\mu\nu} + \hat{\Sigma}_2(\phi, \hat{X})\phi_\mu\phi_\nu; \chi\big),
\end{split}
\end{equation}
with $\hat{\Box}$ and $\hat{\nabla}$ being now defined w.r.t. the Levi-Civita connection of the metric $\hat{g}$, and the functions $\{\hat{B}, \hat{C}, \hat{F}, \hat{\Sigma}_1, \hat{\Sigma}_2\}$ are presented in Appendix \ref{sec:app1}. As we can see, as a result we have obtained an extension of the metric KGB theory. The novelty here is the term proportional to $\hat{C}$, but as it was mentioned before, one can obtain dynamically equivalent theory without the $\hat{B}$-term, with modified $\hat{C}$ and $\hat{F}$. One could carry out an additional conformal transformation of the metric tensor in order to get a non-minimal coupling of the scalar field to the curvature, bringing the action to the form identical to a specific subclass of Horndeski's gravity. 

The theory becomes much easier to analyze in the metric setup. However, owing to the fact that the metric used in action is the invariant metric, it corresponds to an infinite number of metric tensors related to each other through a disformal transformation. The disformal transformation does not seem to be a symmetry of nature, and therefore only one metric can be thought of as describing physical phenomena; alternatively, one could assume that only invariant combinations of the quantities present in the theory should correspond to physical observables, but this again introduces non-uniqueness in the choice of the invariant objects. We have presented a specific choice of the invariant metric and the invariant connection, but equally well one could have chosen:
\begin{equation}\label{eq:inv_met_jor}
    \tilde{g}_{\mu\nu} = \Sigma_1(\phi, X)g_{\mu\nu} + \Sigma_2(\phi, X)\phi_\mu\phi_\nu,
\end{equation}
as an invariant metric; such a choice would effectively transform the action to the Jordan-like frame, in which there is no coupling between the scalar field and the matter part. One could in principle get rid of the anomalous coupling present in \eqref{eq:inv_act_ein} by defining the metric:
\begin{equation}
    \tilde{g}_{\mu\nu} =\hat{\Sigma}_1(\phi, \hat{X})\hat{g}_{\mu\nu} + \hat{\Sigma}_2(\phi, \hat{X})\phi_\mu\phi_\nu,
\end{equation}
identical to \eqref{eq:inv_met_jor}, as can be shown after a simple calculation, and solve it for $\hat{g}_{\mu\nu}$ to get:
\begin{equation}
    \hat{g}_{\mu\nu} = \frac{1}{\hat{\Sigma}_1(\phi, \hat{X}(\tilde{X}))}\tilde{g}_{\mu\nu} - \frac{\hat{\Sigma}_2(\phi, \hat{X}(\tilde{X}))}{\hat{\Sigma}_1(\phi, \hat{X}(\tilde{X}))}\phi_\mu\phi_\nu,
\end{equation}
where $\tilde{X} = \tilde{g}^{\mu\nu}\phi_\mu\phi_\nu$ and the relation $\hat{X}(\tilde{X})$ can be obtained by inverting:
\begin{equation}
    \tilde{X} = \frac{\hat{X}}{\Sigma_1(\phi, \hat{X}) +\Sigma_2(\phi, \hat{X})\hat{X}}.
\end{equation}
In this case, we will end up with a theory in a Jordan-like frame after expressing all the terms present in the action using the metric $\tilde{g}$. For example, one would have to compute $\hat{R}\left(\frac{1}{\hat{\Sigma}_1(\phi, \hat{X}(\tilde{X}))}\tilde{g}_{\mu\nu} - \frac{\hat{\Sigma}_2(\phi, \hat{X}(\tilde{X}))}{\hat{\Sigma}_1(\phi, \hat{X}(\tilde{X}))}\phi_\mu\phi_\nu\right)$, which would result in a plethora of new terms entering the action, effectively taking us away from Horndeski's theory (invariant under disformal transformation of the form $\bar{g}_{\mu\nu} = A(\phi)g_{\mu\nu} + B(\phi)\phi_\mu\phi_\nu$, but not under transformations including the functional dependence on the kinetic term $X$) towards general DHOST theories; in order to see the transformation formulae, the interested Reader is referred to Ref. \cite{achour:2024}. 

\subsubsection*{Example:  Palatini derivative ST theory}

As an example, let us apply the methods of solving the model to the Palatini derivative ST theory, analyzed in Ref. \cite{galtsov:2019}. The gravitational action proposed in that paper is the following:
\begin{equation}
    S_g(g, \Gamma, \phi) = \frac{1}{2\kappa^2}\int d^4x \sqrt{-g}\big((1-\kappa_1 X)g^{\mu\nu}R_{\mu\nu}(\Gamma) - \kappa^2 \phi^\mu\phi^\nu R_{\mu\nu}(\Gamma) - X\big),
\end{equation}
with $\kappa_1, \kappa_2$ being some constants and the matter part neglected. One can make the following identifications: $A_1(\phi, X) =1-\kappa_1 X$, $A_2(\phi, X) =-\kappa_2$, $B(\phi, X) = C(\phi, X) = E_1(\phi, X) = E_2(\phi, X) = E_3(\phi, X) = 0$, $F(\phi, X) = -X$. Then, the invariant metric reads as:
\begin{equation}
    \hat{g}_{\mu\nu} = \sqrt{(1-\kappa_1 X)\big(1-(\kappa_1+\kappa_2) X\big)}\Big(g_{\mu\nu} + \frac{\kappa_2}{1-(\kappa_1+\kappa_2) X}\phi_\mu\phi_\nu\Big).
\end{equation}
Using the transformation formulae from Appendix \ref{sec:app1}, we arrive at the following metric, dynamically equivalent frame:
\begin{equation}
    S_g(\hat{g}, \phi) = \frac{1}{2\kappa^2}\int d^4x \sqrt{-g}\left(\hat{R}(\hat{g}) - \frac{\hat{X}}{1-(\kappa_1 + \kappa_2)X(\hat{X})}\right),
\end{equation}
where the relation between the kinetic terms is given by:
\begin{equation}
    \hat{X} = \frac{X}{(1-\kappa_1 X)^{3/2}\sqrt{1 - (\kappa_1 + \kappa_2)X}}.
\end{equation}

\subsection{Field equation for the scalar field}

An analysis of the field equation for the scalar field was nearly impossible in the general setup given by the action \eqref{eq:action}. However, in the metric approach, the field equations are much simpler:
\begin{equation}
\begin{split}
    & \big(2\hb_{,\hx}(\phi, \hx) - \hc(\phi, \hx)\big)\Big(R^\lambda_{\:\mu\nu\alpha}\phi_\lambda\hat{\phi}^\alpha \hg^{\mu\nu} - (\hat{\Box}\phi)^2 -2\hp^\alpha\hp^\beta\hat{\nabla}_\alpha\phi_\beta\hat{\Box}\phi + 2\hp^\alpha\hp^\beta\hat{\nabla}^\lambda\phi_\alpha \hat{\nabla}_\lambda\phi_\beta + \hat{\nabla}^\alpha\phi^\beta \hat{\nabla}_\alpha\phi_\beta\Big) \\
    &\quad\quad + \hat{\Box}\phi \big(2\hb_{,\phi}(\phi, \hx) -2\hb_{,\phi\hx}(\phi, \hx)\hx+2\hc_{, \phi}(\phi, \hx)\hx - 2\hf_{,\hx}(\phi, \hx)\big) \\
    &\quad\quad + \hp^\alpha\hp^\beta\hat{\nabla}_\alpha\phi_\beta \big(4\hb_{,\phi\hx}(\phi, \hx)+2\hc_{,\phi}(\phi, \hx)+2\hc_{,\hx\phi}(\phi, \hx)\hx-4\hf_{,\hx\hx}(\phi, \hx)\big)\\
    & \quad\quad + \hb_{,\phi\phi}(\phi, \hx)\hx + \hc_{,\phi\phi}(\phi, \hx)\hx^2-2\hf_{,\hx\phi}(\phi, \hx)\hx+\hf_{,\phi}(\phi, \hx) = \kappa^2\left(\hat{T} \frac{\bar{\hat{\Sigma}}_{1, \phi}(\phi, \hx)}{\bar{\hat{\Sigma}}_{1}(\phi, \hx)} + \frac{\bar{\hat{\Sigma}}_{2, \phi}(\phi, \hx)}{\bar{\hat{\Sigma}}_{1}(\phi, \hx)}\hat{T}_{\mu\nu}\hp^\mu\hp^\nu\right) \\
    &\quad\quad\quad\quad\quad\quad - 2\kappa^2\hat{\nabla}_\alpha\Bigg(\hat{T}_{\mu\nu}\left(\frac{\bar{\hat{\Sigma}}_{1, \hx}(\phi, \hx)}{\bar{\hat{\Sigma}}_{1}(\phi, \hx)}\hg^{\mu\nu}\hp^\alpha + \frac{\bar{\hat{\Sigma}}_{2,\hx}(\phi, \hx)}{\bar{\hat{\Sigma}}_{1}(\phi, \hx)}\hp^\alpha\hp^\mu\hp^\nu + \frac{\bar{\hat{\Sigma}}_{2}(\phi, \hx)}{\bar{\hat{\Sigma}}_{1}(\phi, \hx)}\hg^{\alpha (\mu}\hp^{\nu)}\right)\Bigg).
\end{split}
\end{equation}
where $\bar{\hat{\Sigma}}_{1}(\phi, \hx) = 1 / \hat{\Sigma}_{1}(\phi, \hx)$ and $\bar{\hat{\Sigma}}_{1}(\phi, \hx) = -\hat{\Sigma}_{2}(\phi, \hx) / (\hat{\Sigma}_{1}(\phi, \hx)\big(\hat{\Sigma}_{1}(\phi, \hx) + \hat{\Sigma}_{2}(\phi, \hx)\hx\big))$, and $\hat{T}_{\mu\nu} = \frac{-2}{\sqrt{-\hat{g}}}\frac{\delta(\sqrt{-\hat{g}} \mathcal{L}_m)}{\delta \hg^{\mu\nu}}$, $\mathcal{L}_m$ being the matter Lagrangian. As one can see, there arises a highly non-trivial modification to the r.h.s. of the field equation for the scalar field. In the case when $\bar{\hat{\Sigma}}_{1}(\phi, \hx) = \bar{\hat{\Sigma}}_{1}(\phi)$ and $\bar{\hat{\Sigma}}_{2}(\phi, \hx) = 0$, we recover the well-known formula for the ST theories is the Wagoner parametrization, i.e., on the r.h.s we will have only $\kappa^2 \hat{T}\frac{\partial}{\partial\phi} \log \bar{\hat{\Sigma}}_{1}(\phi, \hx)$. Interestingly, the equation simplifies significantly in the case when $\hc(\phi, \hx) = 2\hb_{,\hx}(\phi, \hx)$. This is an obvious consequence of the fact that, when integrating by parts, the term $\hb \hat{\Box}\phi$ can be written as $\hb \hat{\Box}\phi = \text{div} - \hb_{,\phi}\hx-2\hb_{,\hx}\hp^\alpha\hp^\beta\hat{\nabla}_\alpha \phi_\beta$, where $\text{div}$ denotes a total divergence that can be omitted. 

In the case of vanishing matter Lagrangian, if an additional condition is satisfied:
\begin{equation}
    F(\phi, \hx) = \int^{\hx}_{\hx_0} \big(\hb_{,\phi}(\phi, \hx') + \hb_{,\phi\hx'}(\phi, \hx')\hx'\big) d\hx' + \mathcal{V}(\phi)=\hb_{,\phi}(\phi, \hx)\hx + \mathcal{V}(\phi) + \mathcal{C},
\end{equation}
where $\mathcal{V}(\phi)$ is an arbitrary function of the scalar field and $\mathcal{C}$ is an integration constant, the scalar field equation of motion turns out to be $\mathcal{V}'(\phi) = 0$, with $\phi = \phi_0$ being a solution of this algebraic equation, leading effectively to GR with cosmological constant: $\Lambda = -\frac{1}{2}\mathcal{V}(\phi_0)$. However, even if this requirement is fulfilled in the case of non-vanishing matter part of the action, then the scalar field will in general retain its dynamical nature due to the non-trivial, kinetic coupling to matter fields. 

\section{Cosmology in the disformally invariant Palatini theory}\label{sec:5}

The purpose of this section will be to focus on one particular simple model, analyzed in the cosmological context, in order to present how the change of the disformal frame might be used for a simplification of calculations. The main problem with the proposed full theory is the abundance of extra functions of the scalar field and the kinetic term, but the characteristic feature is that, in the Einstein-like frame, there appears a non-trivial coupling of the kinetic term to the matter field. The presence of such coupling could have an impact on the well-posedness of the theory, but this issue remains so far unexplored in the context of Palatini theory.

One of the main difficulties in such calculations is the problem of inversion of the functional dependence between the initial and the invariant kinetic terms. For most of the choices, the relation $\hat{X}(X)$ is complicated, effectively making it impossible to find an analytical formula for $X(\hat{X})$. However, we will focus on a class of theories for which such an inversion is possible for all values of $X$.

Let us consider the following general action functional in the Palatini approach:
\begin{equation}\label{eq:gen-act}
    \begin{split}
        S(g, \Gamma, \phi; \chi) = & \frac{1}{2\kappa^2}\int d^4x \sqrt{-g}\Big(\zeta_1 \phi^{p_1} X^{m_1} g^{\mu\nu}R_{\mu\nu}(\Gamma) + \zeta_2 \phi^{p_2} X^{m_2} \phi^\mu\phi^\nu R_{\mu\nu}(\Gamma) + \zeta_3 f_1(\phi, X) \phi^\mu\phi^\nu \nabla_\mu\nabla_\nu\phi\\
        & + \zeta_4 f_2(\phi, X) X g_{\mu\nu}Q_{\alpha}^{\:\mu\nu}\phi^\alpha + \zeta_5 f_3(\phi, X) Q_\alpha^{\:\mu\nu}\phi_\mu\phi_\nu\phi^\alpha - \Lambda\Big) + S_m(g_{\mu\nu}; \chi),
    \end{split}
\end{equation}
where $p_1, p_2, m_1, m_2, \zeta_1, \zeta_2. \zeta_3, \zeta_4, \zeta_5, \Lambda$ are constant, and $f_1, f_2, f_3$ are arbitrary functions of the scalar field the kinetic term. This particular form of the action was chosen to provide a simple metric counterpart if some conditions are satisfied. At the same time, we make it sufficiently general to include various simple couplings of the scalar field to the curvature and the Ricci tensor. In its full form, the dynamically equivalent metric theory will be complicated; however, if one assumes the following:
\begin{equation}
    m_2 = m_1 -1, \quad \zeta_3 = 24(m_1-1)\zeta_4, \quad \zeta_5 = 4(-4+3m_1)\zeta_4,
\end{equation}
together with:
\begin{equation}
    f_1(\phi, X) = f_2(\phi, X) = f_3(\phi, X) = \mathcal{F}(\phi, X),
\end{equation}
then the metric action, expressed in terms of the invariant metric will read as:
\begin{equation}
    S(\hat{g}, \phi; \chi) = \frac{1}{2\kappa^2}\int d^4x \sqrt{-\hg}\big(\hat{R}(\hg) + \hat{F}(\phi, \hat{X})\big) + S_m\left(\hat{\Sigma}_1(\phi,\hx)\hg_{\mu\nu} + \hat{\Sigma}_2(\phi,\hx)\phi_\mu\phi_\nu; \chi\right),
\end{equation}
with:
\begin{align}
     & \hat{F}(\phi, \hat{X}) = a_1(\phi)\mathcal{F}(\phi, \hx)\hat{X}^2 + a_2(\phi)\mathcal{F}(\phi, \hx)^2\hat{X}^3 - \Lambda a_3(\phi)\hx^{\frac{2m_1}{m_1-1}},\label{eq:f-func} \\
     & \hat{\Sigma}_1(\phi,\hx) = \Big((\zeta_1 \phi^{p_1})^{1+ 2m_1} (\zeta_1 \phi^{p_1} + \zeta_2\phi^{p_2})^{1-2m_1}\Big)^{\frac{1}{2(m_1 - 1)}} \hx^{\frac{m_1}{m_1 - 1}}, \\
     & \hat{\Sigma}_2(\phi,\hx) = \zeta_2 \phi^{p_2}(\zeta_1 \phi^{p_1})^\frac{3}{2(m_1 - 1)}(\zeta_1 \phi^{p_1} + \zeta_2^{p_2})^{-1+\frac{1}{2-2m_1}} \hx^{\frac{1}{m_1 - 1}}, \\
     & \hx = \frac{X^{1 - m_1}\sqrt{\zeta_1\phi^{p_1} + \zeta_2\phi^{p_2}}}{\zeta_1^{3/2} \phi^{\frac{3p_1}{2}}},\label{eq:x-transf}
\end{align}
where $\mathcal{F}(\phi, \hx)$ is understood as $\mathcal{F}(\phi, X(\hx))$. In the last line, we expressed the new kinetic term as a function of the old one and the scalar field; this relation can be easily inverted. Moreover, we also introduced:
\begin{align}
    & a_1(\phi) = 3\sqrt{\zeta_1}\zeta_4\frac{\phi^{-1 + \frac{p_1}{2}}}{2(\zeta_1\phi^{p_1}+\zeta_2 \phi^{p_2})^{5/2}} \Big(\zeta_2 p_2\phi^{p_2}(\zeta_2(4m_1-5)\phi^{p_2} - 2\zeta_1(2m_1-3)\phi^{p_1}) +p_1(8(m_1-1)\zeta_1^2\phi^{2p_1} \nonumber \\
    & \quad\quad\quad - 3(4m_1-5)\zeta_2^2\phi^{2p_2} + 4(m_1-1)\zeta_1\zeta_2 \phi^{p_1+p_2})\Big), \\
    & a_2(\phi) = \frac{6\zeta_1^3\zeta_4^2\phi^{3p_1}}{(\zeta_1\phi^{p_1} + \zeta_2 \phi^{p_2})^3}, \\
    & a_3(\phi) = (\zeta_1\phi^{p_1})^{\frac{3(m_1 + 1)}{2(m_1 - 1)}}(\zeta_1\phi^{p_1} + \zeta_2 \phi^{p_2})^{\frac{3m_1 -1}{-2(m_1 - 1)}}.
\end{align}

 As we can see, the exponents of the kinetic terms in the $\hat{F}$ function do not depend on the $m_1$ parameter. Another interesting feature is that, even though there might be no potential or kinetic term for the scalar field in the original action (i.e., $f_i(\phi, X) = 1$ in \eqref{eq:gen-act}), it arises on-shell as a result of integrating out the auxiliary connection. The simplification of the gravitational part of the action comes at the cost of introducing non-trivial couplings to the matter part, leading to complicated field equations for the scalar field involving derivatives of the stress-energy tensor. 

Let us observe the following fact: the gravitational part of the action extends formally the $k$-essence type of theories \cite{picon:2000, rendall:2006}, i.e., to a class of theories with a dynamical scalar field (or fields) with the kinetic term entering the action in a non-trivial manner, usually in the following form: $\hat{F}(\phi, \hx) = f_1(\phi)f_2(\hx)$, where $f_1, f_2$ are some functions of their respective arguments \cite{picon:1999, scherrer:2000, chiba:2000}. These theories arise naturally in string theories at the level of effective action \cite{picon:2000}. Models bearing greater resemblance to the one presented here have been analyzed in Ref. \cite{chakraborty:2019, shi:2021}, where the $\hat{F}(\phi, \hx)$ function is given as $\hat{F}(\phi, \hx) = a_1(\phi)\hx + a_2(\phi)\hx^2 + V(\phi)$, with $a_1, a_2, V$ being some functions of the scalar field. 

The theory itself still displays considerable level of complicacy, therefore we will make a couple of simplifying assumptions: we use the standard set of spherical coordinates with cosmic time $t$, assuming that all relevant dynamical quantities will depend on it (i.e., $\phi = \phi(t))$; we also put $\kappa = 1$ and neglect the pressure (i.e., we consider only the dust). Moreover, we will assume that the metric $\hat{g}$ is the Friedmann-Robertson-Lema\^{i}tre-Walker metric of a spatially flat, expanding Universe: $\hat{g}_{\mu\nu} = \text{diag}(-1, \hat{a}^2(t),  \hat{a}^2(t) r^2,  \hat{a}^2(t) r^2 \sin^2\theta)$ in the standard spherical coordinates.  The equations of motion will hence read as follows:
\begin{align}
    & \hat{G}_{\mu\nu}(\hg)= \hat{T}^{(m)}_{\mu\nu} + \hat{T}^{(\phi)}_{\mu\nu}, \\
    & \hat{F}_{,\phi}(\phi, \hx) - 2\hat{F}_{,\phi\hx}(\phi, \hx)\hx - 4\hat{F}_{,\hx\hx}(\phi, \hx)\hp^\mu\hp^\nu\hat{\nabla}_\mu\phi_\nu - 2\hat{F}_{,\hx}(\phi, \hx)\hat{\Box}\phi = \nonumber \\
    &\quad\quad\quad - \hat{\rho}\left(\frac{\bar{\hat{\Sigma}}_{1, \phi}(\phi, \hx)}{\bar{\hat{\Sigma}}_{1}(\phi, \hx)} + \hx \frac{\bar{\hat{\Sigma}}_{2, \phi}(\phi, \hx)}{\bar{\hat{\Sigma}}_{1}(\phi, \hx)}\right) \nonumber \\
    & \quad\quad\quad -2\hat{\nabla}_\alpha\left(-\hat{\rho}\left(\frac{\bar{\hat{\Sigma}}_{1, \hx}(\phi, \hx)}{\bar{\hat{\Sigma}}_{1}(\phi, \hx)}\hp^\alpha + \hx \frac{\bar{\hat{\Sigma}}_{2, \hx}(\phi, \hx)}{\bar{\hat{\Sigma}}_{1}(\phi, \hx)}\hp^\alpha + \dot{\hp}\frac{\bar{\hat{\Sigma}}_{2}(\phi, \hx)}{\bar{\hat{\Sigma}}_{1}(\phi, \hx)} \delta^\alpha_{0} \right)\right),
\end{align}
where $\dot{\phi} = \frac{d\phi}{dt}$, $\hat{u}_\mu = (-1, 0,0,0)$, $\hat{T}^{(m)}_{\mu\nu}=\hat{\rho}\hat{u}_\mu\hat{u}_\mu$ is the energy-momentum tensor for the dust, $\hat{T}^{(\phi)}_{\mu\nu} = -\hat{F}_{,\hx}(\phi, \hx)\phi_\mu\phi_\nu + \frac{1}{2}\hat{F}(\phi, \hx)\hg_{\mu\nu}$ is the energy-momentum tensor of the scalar field, which will take a more familiar form if we write it as: 
\begin{equation}
    \hat{T}^{(\phi)}_{\mu\nu} = (\hat{\rho}_\phi + \hat{p}_\phi)\hat{u}^{(\phi)}_\mu \hat{u}^{(\phi)}_\nu + \hat{p}_\phi \hg_{\mu\nu},
\end{equation}
with:
\begin{equation}
    \hat{p}_\phi = \frac{1}{2}\hat{F}(\phi, \hx), \quad \hat{\rho}_\phi = -\frac{1}{2}\hat{F}(\phi, \hx) + \hat{F}_{,\hx}(\phi, \hx)\hx, \quad \hat{u}^{(\phi)}_\mu = \frac{\phi_\mu}{\sqrt{-\hx}}.
\end{equation}

The ansatz concerning the form of the metric is applicable to the Einstein-like frame only; in general, upon a change of the frame to the initial one, the original metric $g$ might lose its FRLW form. Nevertheless, we assume that in the other frame we keep the same set of coordinates, so that the time derivative stays the same, i.e., one does not transform $\dot{\phi}$. On the other hand, it was necessary to denote the scale factor as an object dependent on the current frame by writing $\hat{a}$; if one changes the disformal frame, in general, the multiplicative factor responsible for controlling the cosmic expansion will also change. After making this ansatz, the relevant field equations will be (we omit the time dependence of the dynamical quantities for a more concise presentation):
\begin{align}
    & 3\hat{H}^2 = -\frac{1}{2}\hat{F}(\phi,\hx)+\hat{F}_{,\hx}(\phi,\hx)\hx + \hat{\rho}\equiv \hat{\rho}_\phi + \hat{\rho}, \label{eq:friedmann-1} \\
    & 2\dot{\hat{H}} + 3\hat{H}^2 = \frac{1}{2}\hat{F}(\phi,\hx) + \hat{p}\equiv \hat{p}_\phi  +\hat{p},\label{eq:friedmann-2} \\
    & \left(2\hat{F}_{,\hx}(\phi,\hx)-4\hat{F}_{,\hx\hx}(\phi,\hx)\hx\right)\ddot{\phi} + 6H\hat{F}_{,\hx}(\phi,\hx)\dot{\phi} + \hat{F}_{,\phi}(\phi,\hx)-2\hat{F}_{,\hx\phi}(\phi,\hx)\hx = \nonumber \\
    &\quad\quad\quad - \hat{\rho}\left(\frac{\bar{\hat{\Sigma}}_{1, \phi}(\phi, \hx)}{\bar{\hat{\Sigma}}_{1}(\phi, \hx)} + \hx \frac{\bar{\hat{\Sigma}}_{2, \phi}(\phi, \hx)}{\bar{\hat{\Sigma}}_{1}(\phi, \hx)}\right) \nonumber \\
    & \quad\quad\quad -2\hat{\nabla}_\alpha\left(-\hat{\rho}\left(\frac{\bar{\hat{\Sigma}}_{1, \hx}(\phi, \hx)}{\bar{\hat{\Sigma}}_{1}(\phi, \hx)}\hp^\alpha + \hx \frac{\bar{\hat{\Sigma}}_{2, \hx}(\phi, \hx)}{\bar{\hat{\Sigma}}_{1}(\phi, \hx)}\hp^\alpha + \dot{\phi}\frac{\bar{\hat{\Sigma}}_{2}(\phi, \hx)}{\bar{\hat{\Sigma}}_{1}(\phi, \hx)} \delta^\alpha_{0} \right)\right)\label{eq:friedmann-3}
\end{align}
where $(\dot{\:\:})$ denotes differentiating w.r.t. the cosmic time $t$, while $\hat{H}=\dot{\hat{a}}/\hat{a}$ (in general, the hat denotes quantities corresponding to the Einstein-like frame). 

We will now aim at eliminating those metric versions of the theory equivalent to \eqref{eq:gen-act} which display pathologies or are dynamically non-viable. We will divide our considerations into two parts: theories with a present energy density $\rho$ and theories in the vacuum. 

\subsection{Vanishing energy density}

Let us notice that the transformation \eqref{eq:x-transf} does not allow for the change of the sign of the kinetic term $X$. Assuming that $\phi > 0$ and $\zeta_1 > 0$, the sign of $X$ will be the same as the sign of $\hx$; in the initial metric setup corresponding to an Einstein-like frame we have $\hat{X} = -\dot{\phi}^2 \leq 0 $, and so one should get $X \leq 0$, whatever it might be in terms of the derivatives of the scalar field. For this reason, $m_1$ has to be a number of the form $m_1 = \pm n\pm\frac{a}{b}$, $n,a, b\in\mathbb{N}$, $a < b$ with $a, n$ even, and $b$ odd. On the other hand, the term containing the cosmological constant in \eqref{eq:f-func} features the expression $\hx^\frac{2m_1}{m_1 - 1} =\hx^{2 +  \frac{2}{m_1 - 1}}$, which is a power of a negative quantity; the exponent is well-defined for all values of $n, a, b$ satisfying the condition given above. 

Below, we write the three independent equations \eqref{eq:friedmann-1}, \eqref{eq:friedmann-2}, and \eqref{eq:friedmann-3} with an additional simplifying assumption concerning the exponent of the scalar field: $p_1 = p_2 = 0$. Also, let us assume that the $\mathcal{F}$ function has the form $\mathcal{F}(\hx) =  c_q \hx^{-q}$:
\begin{align}
    3\hat{H}^2 & = \frac{3(5 -4q)c_q^2\zeta_1^3\zeta_4^2}{(\zeta_1 + \zeta_2)^3}\hx^{3 -2q} -\frac{(1+3m_1)\Lambda}{2(m_1-1)} \zeta_1^{\frac{3m_1}{2(m_1-1)}}(\zeta_1 + \zeta_2)^{\frac{3m_1 -1}{2(1 - m_1)}}\hx^\frac{2m_1}{m_1-1} \equiv \hat{\rho}_\phi, \\
    3\hat{H}^2 + 2\dot{\hat{H}} & = -\frac{\Lambda}{2} \zeta_1^\frac{3m_1}{2(m_1-1)} (\zeta_1+\zeta_2)^\frac{3m_1 - 1}{2(1 - m_1)}\hx^{\frac{2m_1}{m_1-1}} - \frac{3c_q^2\zeta_1^3\zeta_4^2}{(\zeta_1 + \zeta_2)^3}\hx^{3 - 2q}\equiv \hat{p}_\phi, \\
    -(5-4q)\ddot{\phi}-3\hat{H}\dot{\phi} & = \Lambda \frac{m_1\hx^{\frac{1+m_1 - 2q+2m_1 q}{m_1 -1}-2}\big(3(m_1-1)\hat{H} \dot{\phi} + (1 + 3m_1)\ddot{\phi}\big)}{3\zeta_4^2(m_1-1)^2(3+2q)}\left( \frac{\zeta_1}{\zeta_1 + \zeta_2}\right)^{\frac{7-3m_1}{2(m_1-1)}} .
\end{align}
At early times, the cosmological constant should not play a significant role, therefore one can ignore it. Then, the system will be of the form:
\begin{align}
    3\hat{H}^2 & = \frac{3(5 -4q)\zeta_1^3\zeta_4^2}{(\zeta_1 + \zeta_2)^3}\hx^{3 - 2q}, \\
    3\hat{H}^2 + 2\dot{\hat{H}} & = - \frac{3\zeta_1^3\zeta_4^2}{(\zeta_1 + \zeta_2)^3}\hx^{3 - 2q}, \\
    -(5-4q)\ddot{\phi}-3\hat{H}\dot{\phi} & = 0,
\end{align}
which can be easily solved:
\begin{align}
    \hat{a}(t) & = \left(\frac{3(3-2q)\left(\mathcal{C}_1^{3-2q}\sqrt{-\frac{\zeta_1^3\zeta_4^2(5-4q)}{(\zeta_1 + \zeta_2)^3}}t + \mathcal{C}_2\right)}{5-4q}\right)^\frac{5-4q}{9-6q}, \\
    \phi(t) & = \mathcal{C}_1\int^t \frac{dt'}{\hat{a}(t')^\frac{3}{5-4q}}\sim \left(\mathcal{C}_1^{3-2q}\sqrt{-\frac{\zeta_1^3\zeta_4^2(5-4q)}{(\zeta_1 + \zeta_2)^3}}t + \mathcal{C}_2\right)^\frac{2(1-q)}{3-2q}.
\end{align}
If $1 < q < \frac{3}{2}$, the scalar field's value will go down to zero as the times increases.
Moreover, in order for the function to make any sense, one requires $\frac{5-4q}{\zeta_1+\zeta_2} < 0$. As it is obvious from the form of the function, the model does support exponential inflation, but rather can mimic the rate of the cosmic expansion for various types of energy with the exception of dust.

\subsection{Non-vanishing energy density}
The presence of matter will greatly complicate the theory due to non-trivial couplings of the components of energy-momentum tensor to the scalar field functions. 
Let us now consider the case when $p_1 = p_2 = 0$, $m_1 = 2$, $\mathcal{F}(\phi, \hat{X}) = c_q \hx^{-q}$. Then, the relevant equations will take the following form:

\begin{align}
  &  3\hat{H}^2  = \hat{\rho} - \Lambda \frac{7\zeta_1^{9/2}}{2(\zeta_1 + \zeta_1)^{5/2}}\hx^4 + \frac{3c_q^2\zeta_1^3\zeta_4^2(5-4q)}{(\zeta_1 + \zeta_1)^3}\hx^{3-2q}, \\
  &  3\hat{H}^2 + 2\dot{\hat{H}} = -\Lambda \frac{\zeta_1^{9/2}}{2(\zeta_1 + \zeta_1)^{5/2}}\hx^4 - \frac{3c_q^2\zeta_1^3\zeta_4^2}{(\zeta_1 + \zeta_1)^3}\hx^{3-2q},
     \\
  &   \frac{\zeta_1^3\hx^2}{(\zeta_1 + \zeta_1)^3}\Bigg[\left(-56\zeta_1^{3/2}(\zeta_1+\zeta_1)^{1/2}\Lambda \hx + 12\zeta_4^2(15-22q+8q^2)\hx^{-2q}\right)\ddot{\phi} \nonumber \\
  &  + \left(-24\zeta_1^{3/2}(\zeta_1+\zeta_1)^{1/2} \Lambda \hx + 36\zeta_4^2(3-2q)\hx^{-2q}\right)\hat{H}\dot{\phi}\Bigg] = -\frac{4\left[3(\zeta_1 - \zeta_1)\hat{H} \rho \dot{\phi} - \zeta_1 \dot{\rho}\dot{\phi} + \zeta_1 \rho \ddot{\phi}\right]}{(\zeta_1 + \zeta_1)\hx^2}.
\end{align}
We will carry out a dynamical system analysis to assess the qualitative behaviour of this system for a couple of values of the parameter $q$.

\subsubsection{Case I: $q = 2$}

We choose the following dimensionless variables:

$$x_1 = \frac{\sqrt{3}c_q \zeta_1^{3/2}\zeta_4}{(\zeta_1 + \zeta_2)^{3/2} \hat{H} \dot{\phi}}, \quad x_2 = \frac{\sqrt{\frac{7}{6}}\zeta_1^{9/4} \sqrt{\Lambda} \dot{\phi}^4}{(\zeta_1 + \zeta_2)^{5/4} \hat{H}},$$

making the Friedmann equation:

\begin{equation}
    1 = \Omega_m + x_1^2 - x_2^2\equiv \Omega_m + \Omega_\phi, \quad \Omega_m = \frac{\hat{\rho}}{3\hat{H}^2}, \quad \Omega_\phi = x_1^2 - x_2^2, 
\end{equation}
with the condition $0 \leq \Omega_m \leq 1 \Rightarrow 0\leq x_1^2 - x_2^2 \leq 1 $. The deceleration parameter becomes:

\begin{equation}
    q =-1-\frac{\dot{\hat{H}}}{\hat{H}} =\frac{1}{2} - \frac{x_1^2}{2} - \frac{3x_2^2}{14}.
\end{equation}
In this case, we get a two-dimensional system written in terms of the dimensionless variables as follows:
\begin{align}\label{eq:system1}
   & \frac{1}{\hat{H}}\frac{\text{d} x_1}{\text{d} t}\equiv \frac{\text{d} x_1}{\text{d}\hat{N}} = \frac{x_1\Big[\zeta_1\Big(21 + 28x_1^4 - 306x_2^2 + 33x_2^4+x_1^2(-49 + 89x_2^2)\Big) + \zeta_2\Big(7x_1^4 + x_1^2(-49 + 31x_2^2) + 6(7 - 11x_2^2+2x_2^4)\Big)\Big]}{14\Big(\zeta_1(-1 + 4x_1^2 + 11x_2^2) + \zeta_2(x_1^2+4x_2^2)\Big)},\nonumber \\
    & \frac{1}{\hat{H}}\frac{\text{d} x_2}{\text{d} t}\equiv \frac{\text{d} x_2}{\text{d}\hat{N}}  = \frac{x_2 \Big[\zeta_1\Big(21 + 54x_2^2 + 28x_1^4 + 33x_2^4+x_1^2(-259 + 89x_2^2)\Big) + \zeta_2\Big(7x_1^4 + x_1^2(91 + 31x_2^2) +  12(-14 -13x_2^2 + x_2^4)\Big)\Big]}{14\Big(\zeta_1 (-1 + 4x_1^2 + 11x_2^2) + \zeta_2(x_1^2 + 4x_2^2)\Big)}.
\end{align}
with the redefinition of the independent time variable in the Einstein frame $\frac{1}{\hat{H}}\frac{\text{d}}{\text{d} t} = \frac{\text{d}}{\text{d} \hat{N}}$. We are now aiming at finding the stationary points of the system $x'_i = f_i(x_1, x_2)$, i.e., values of $x_1, x_2$ s.t. $\frac{\text{d} x_i}{\text{d}\hat{N}}\equiv x_i' = 0$ for $i = 1,2$, and assessing their stability. In order to do that, we linearize the system around the critical points and investigate the eigenvalues of the Jacobian $J = \frac{\partial f_i}{\partial x_j}$. In Table \ref{tab:system1}, we first list all the real stationary points of the system.

Let us observe that the system features values of $x_1, x_2$, for which the denominator vanishes, leading to a non-smooth dynamical system that cannot be analyzed using the standard techniques. Investigating the system near the singular regions requires separate treatment, see Ref. \cite{saavedra:2001, kunze:2000}.

\begin{table}[]
\begin{tabular}{c | c | c | c | c}
 & $x_1$ &  $x_2$ & $\Omega_\phi$ & Existence \\
 \hline\hline
$A_{\pm}$ & $0$ & $\pm\sqrt{\frac{7(8\zeta_2 - \zeta_1)}{11\zeta_1 + 4\zeta_2}}$ & $-\frac{7(8\zeta_2 - \zeta_1)}{11\zeta_1 + 4\zeta_2}$ &  None \\
$B_{\pm, \pm}$ & $\pm \sqrt{\frac{12}{5}}$  & $\pm\sqrt{\frac{7}{5}}$ & $1$ & All $\zeta_1, \zeta_2$ \\
$C_{\pm}$ & $\pm 1$ & $0$  & 1 & All $\zeta_1, \zeta_2$ \\
$D$ & $0$ & $0$ & $0$ & All $\zeta_1, \zeta_2$ \\
$E_{\pm}$ & $\pm \sqrt{\frac{3(\zeta_1 + 2\zeta_2)}{4\zeta_1 + \zeta_2}}$ & 0 & $\frac{3(\zeta_1 + 2\zeta_2)}{4\zeta_1 + \zeta_2}$ & $-\frac{\zeta_1}{2} < \zeta_2 < \frac{\zeta_1}{5}$
\end{tabular}
\caption{Fixed points of the system \eqref{eq:system1}.}
\label{tab:system1}
\end{table}

Let us now discuss the stability properties of these points:

\begin{itemize}
    \item \textbf{Points $A_{\pm}$}: lie outside of the region.
    \item \textbf{Points $B_{\pm, \pm}$}: source for $ 0 < \zeta_1 < \frac{|\zeta_2|}{3}$, non-hyperbolic for $\zeta_2 = 3\zeta_1$, otherwise a saddle point.
    \item \textbf{Points $C_{\pm}$}: source for $\left(\zeta_2 \leq 0 \land \zeta_1 > -\frac{\zeta_2}{3}\right) \lor \left(\zeta_2 > 0 \land \zeta_1 > 5\zeta_2\right)$, non-hyperbolic for $\zeta_1 = 5\zeta_2$, otherwise a saddle point.
    \item \textbf{Point $D$}: stable for $\zeta_2 >0 \land 0 < \zeta_1 < 8\zeta_2$, source for $\zeta_2 < 0 \land 0 < \zeta_1 < -2\zeta_2$, non-hyperbolic for $\zeta_2 = -\frac{\zeta_1}{2} \lor \zeta_2 = \frac{\zeta_1}{8}$, otherwise a saddle point.
    \item \textbf{Points $E_{\pm}$}: source for $(\zeta_2 < 0 \land -2\zeta_2 < \zeta_1< -3\zeta_2) \lor \left(\zeta_2 > 0 \land \frac{\zeta_2}{3} < \zeta_1 < 5\zeta_2\right)$, non-hyperbolic for $\zeta_2 = -\frac{\zeta_1}{2} \lor \zeta_2 = \frac{\zeta_1}{5}\lor \zeta_2 = 3\zeta_1$, otherwise a saddle.
\end{itemize}

In the case of the points at which the Jacobian $J$ has at least one eigenvalue with zero real part, one cannot apply the usual techniques and must rather resort to the center manifold analysis \cite{glendinning:1994, marsden:1976, carr:1982}. Such an analysis, although straightforward, will not be carried out for the proposed models in this paper, as we will limit our attention to the points that are hyperbolic.

\begin{figure}[h]
\includegraphics[width=8cm]{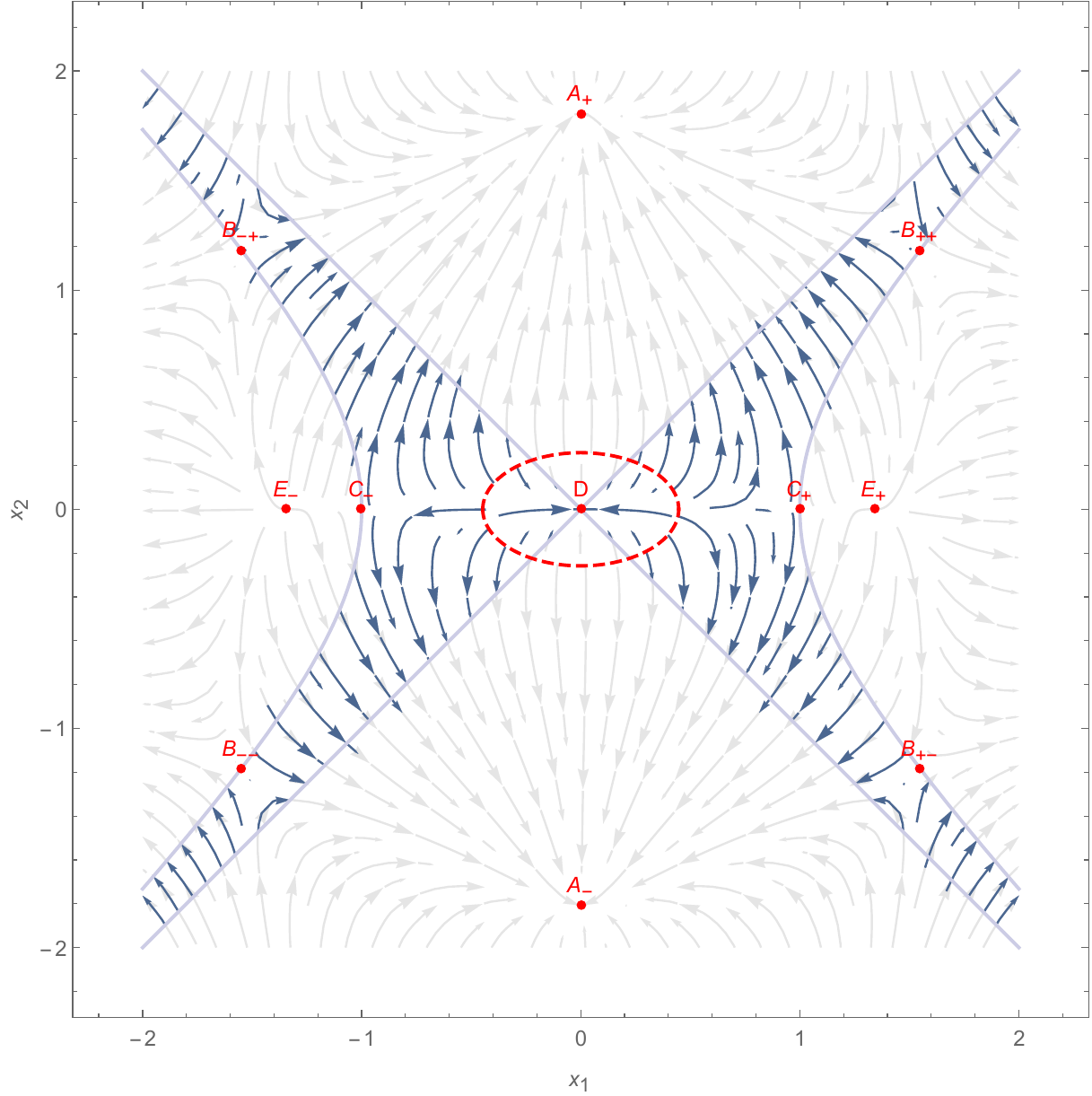}
\caption{Phase space of the system \eqref{eq:system1} with $\zeta_1 = \zeta_2$. The red dashed line represents those values of $x_1, x_2$ for which the system becomes singular. The part of the figure in light blue corresponds to the unphysical region within the phase space.}
\label{fig:phase_space_1}
\centering
\end{figure}

The phase space of the system is presented in Fig \ref{fig:phase_space_1} with $\zeta_1 = \zeta_2$. As one can see, the only stable point for this set of the parameters is the point $D$, for which the energy density related to the scalar field vanishes, resembling a matter-dominated epoch. Indeed, at this point, $q = \frac{1}{2}$, suggesting that $a(t) \sim t^{2/3}$.

Nevertheless, for all possible values of the parameters $\zeta_1, \zeta_2$, there is only one stable point whose analysis can be carried out without invoking the center manifold theorem. Although the points $B_{\pm, \pm}$ correspond to a universe dominated by the scalar field ($\Omega_\phi = 1$ in the hyperbola $x_1 ^2 - x_2^2 = 1$) with the deceleration parameter $q = -1$, the fixed points are unstable and cannot represent the final stage of cosmic evolution. For this reason, the model with $q = -2$ does not seem to offer a viable representation of the dynamics of the scale factor.

\subsubsection{Case II: $q = 4$}

We choose now as our dimensionless variables:

$$x_1 = \frac{\sqrt{11} c_q\zeta_1^{3/2}\zeta_4}{(\zeta_1 + \zeta_2)^{3/2} \hat{H} \dot{\phi}^5}, \quad x_2 = \frac{\sqrt{\frac{7}{6}}\zeta_1^{9/4} \sqrt{\Lambda} \dot{\phi}^4}{(\zeta_1 + \zeta_2)^{5/4} \hat{H}}.$$

The deceleration parameter becomes:
\begin{equation}
    q = \frac{1}{2} - \frac{3x_1^2}{22} - \frac{3x_2^2}{14}.
\end{equation}
Let us notice that only the variable $x_1$ has changed in comparison to the case $q = 2$. The Friedmann equation, as well as the condition for the scalar field dimensionless energy density parameter, are exactly the same, too. With these variables, the relevant equations are now:
{\small
\begin{align}\label{eq:system2}
   & \frac{1}{\hat{H}}\frac{\text{d} x_1}{\text{d} t}\equiv \frac{\text{d} x_1}{\text{d}\hat{N}} = \frac{3 x_1\Big[\zeta_1\Big(112 x_1^4 + x_1^2(-189 + 253x_2^2) + 11(7-198x_2^2 +11x_2^4)\Big) + \zeta_2\Big(35x_1^4 + x_1^2(-805 + 83x_2^2) + 22(35 + x_2^2 + 2x_2^4\Big)\Big]}{154\Big(\zeta_1(-1 + 16x_1^2 + 11x_2^2) + \zeta_2(5x_1^2+4x_2^2)\Big)},\nonumber \\
    & \frac{1}{\hat{H}}\frac{\text{d} x_2}{\text{d} t}\equiv \frac{\text{d} x_2}{\text{d}\hat{N}}  = \frac{3x_2 \Big[\zeta_1\Big(112x_1^4 + 11x_1^2(-49 + 83x_2^2) +11(7+18x_2^2 + 11x_2^4))\Big) + \zeta_2\Big(35x_1^4 + x_1^2(-49 + 83x_2^2) + 44(-14-13x_2^2+x_2^4))\Big)\Big]}{154\Big(\zeta_1(-1 + 16x_1^2 + 11x_2^2) + \zeta_2(5x_1^2+4x_2^2)\Big)}.
\end{align}}

\begin{table}[]
\begin{tabular}{c | c | c | c | c}
 & $x_1$ &  $x_2$ & $\Omega_\phi$ & Existence \\
 \hline\hline
$A_{\pm}$ & $0$ & $\pm\sqrt{\frac{7(8\zeta_2 - \zeta_1)}{11\zeta_1 + 4\zeta_2}}$ & $-\frac{7(8\zeta_2 - \zeta_1)}{11\zeta_1 + 4\zeta_2}$ &  None \\
$B_{\pm, \pm}$ & $\pm \frac{2\sqrt{11}}{3}$  & $\pm\frac{\sqrt{35}}{3}$ & $1$ & All $\zeta_1, \zeta_2$ \\
$C_{\pm}$ & $\pm 1$ & $0$  & 1 & All $\zeta_1, \zeta_2$ \\
$D$ & $0$ & $0$ & $0$ & All $\zeta_1, \zeta_2$ \\
$E_{\pm}$ & $\pm \sqrt{\frac{11(\zeta_1 + 10\zeta_2)}{16\zeta_1 + 5\zeta_2}}$ & 0 & $\frac{11(\zeta_1 + 10\zeta_2)}{16\zeta_1 + 5\zeta_2}$ & $-\frac{\zeta_1}{10} < \zeta_2 < \frac{\zeta_1}{21}$
\end{tabular}
\caption{Fixed points of the system \eqref{eq:system1}.}
\label{tab:system2}
\end{table}

The fixed points of this system are given in Table \ref{tab:system2}. As we can see, the points $A_{\pm}, C_{\pm}, D$ remain unchanged. It does not mean, though, that their stability properties are the same. The detailed analysis is given below:

\begin{itemize}
    \item \textbf{Points $A_{\pm}$}: lie outside of the region.
    \item \textbf{Points $B_{\pm, \pm}$}: stable for $ -3\zeta_1 < \zeta_2 < 3\zeta_1$, non-hyperbolic for $\zeta_2 = 3\zeta_1$, otherwise a saddle point.
    \item \textbf{Points $C_{\pm}$}: source for $\left( \zeta_2 < 0 \land \zeta_1 < -\frac{\zeta_2}{3}\right) \lor (\zeta_2 > 0 \land \zeta_2 < 21\zeta_2)$, non-hyperbolic for $\zeta_2 = \frac{\zeta_1}{21}$, otherwise a saddle.
    \item \textbf{Point $D$}: source for $(\zeta_2 \leq 0 \land \zeta_1 > -10\zeta_2)\lor (\zeta_2 > 0 \land \zeta_1 > 8\zeta_2)$, non-hyperbolic for $\zeta_2 = -\frac{\zeta_1}{10} \lor \zeta_2 = \frac{\zeta_1}{8}$, otherwise a saddle.
    \item \textbf{Points $E_{\pm}$}: stable for $\left(\zeta)_2 < 0\land \zeta_1 < -\frac{5\zeta_2}{16}\right) \lor \left(\zeta_2 > 0\land \zeta_1 < \frac{\zeta_2}{3}\right)$, source for $\left(\zeta_2 \leq 0 \land \left(-\frac{5\zeta_2}{16} < \zeta_1 < -10\zeta_2 \lor \zeta_1 > -11\zeta_2\right)\right) \lor (\zeta_2 >0 \land \zeta_1 > 21 \zeta_2)$, non-hyperbolic for $\zeta_2 = -\frac{\zeta_1}{10} \lor \zeta_2 = \frac{\zeta_1}{21} \lor \zeta_2 = 3\zeta_1$, otherwise a saddle.
\end{itemize}

\begin{figure}[htbp]
    \centering
    \begin{subfigure}[b]{0.45\textwidth}
        \centering
        \includegraphics[width=\textwidth]{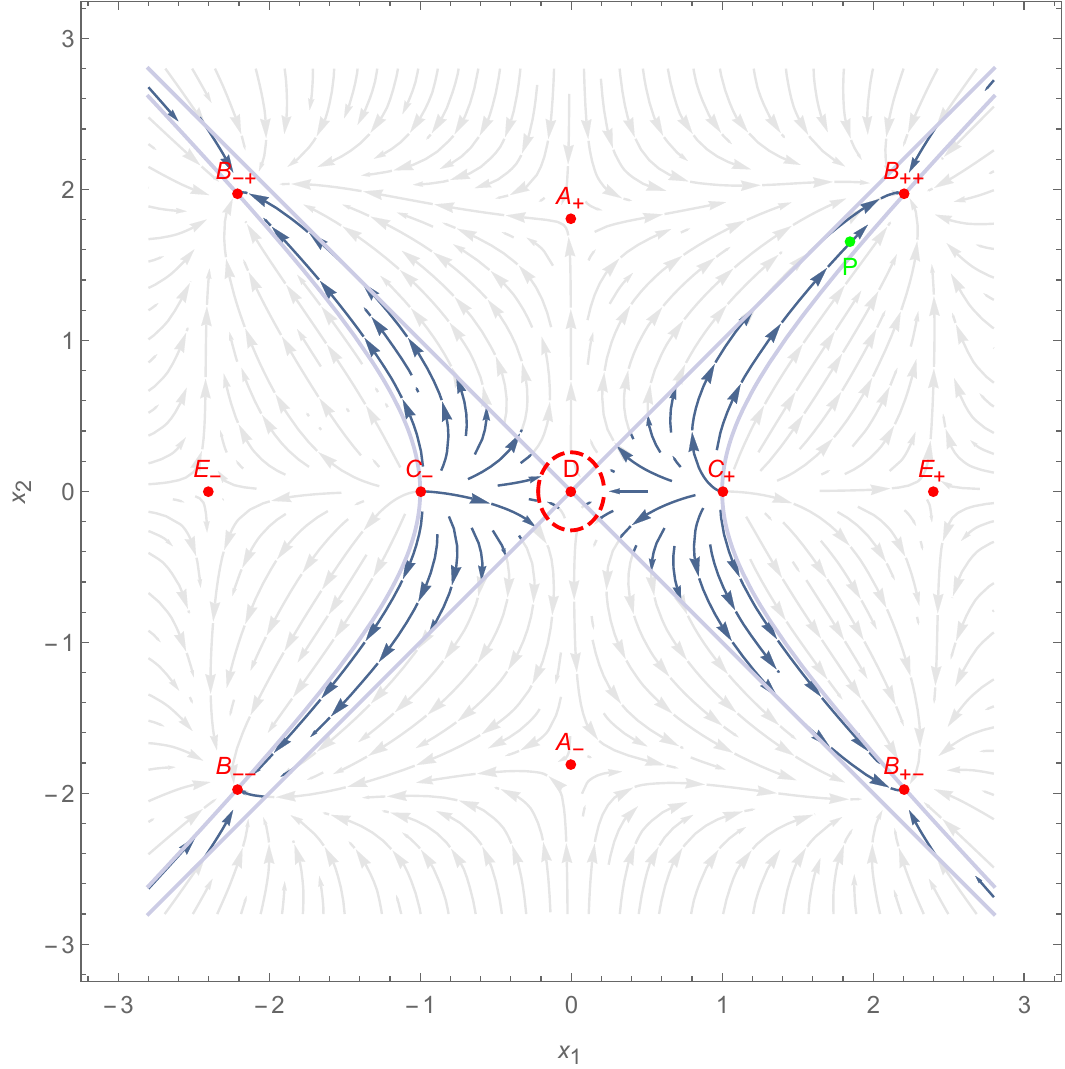}
        \caption{$\zeta_1 = \zeta_2$}
        \label{fig:phase_space_2}
    \end{subfigure}
    \hfill 
    \begin{subfigure}[b]{0.45\textwidth}
        \centering
        \includegraphics[width=\textwidth]{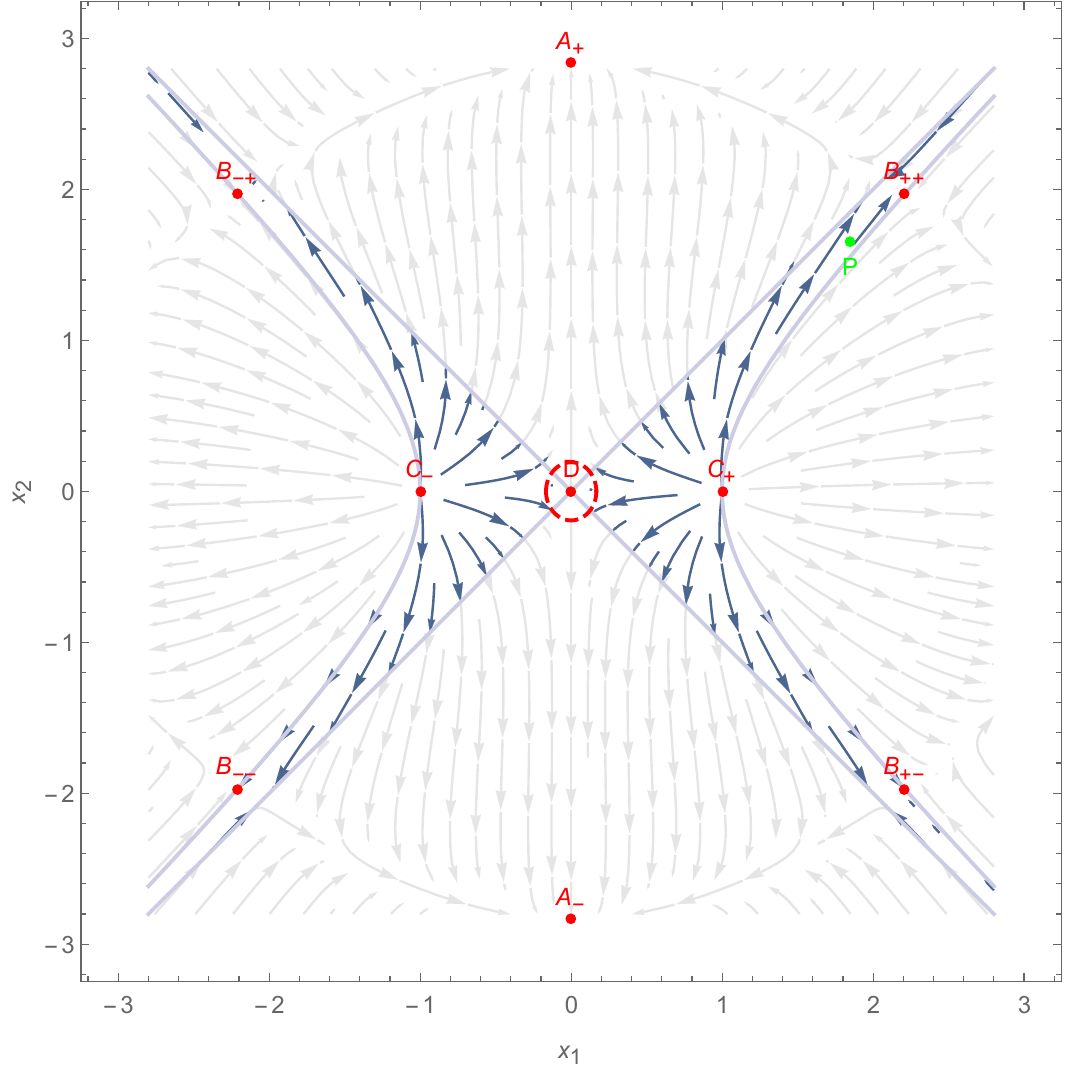}
        \caption{$\zeta_1 = \frac{\zeta_2}{4}$}
        \label{fig:phase_space_3}
    \end{subfigure}
    \caption{Phase space of the system \eqref{eq:system2}. The red dashed line represents those values of $x_1, x_2$ for which the system becomes singular. The green point $P$ represents the possible current state of the universe. The part of the figure in light blue corresponds to the unphysical region within the phase space.}
    \label{fig:both}
\end{figure}

We carry out some analysis for the case $\zeta_1 = \zeta_2$. As it can be seen from the list given above, but also from inspecting Fig \ref{fig:phase_space_2} depicting the phase space trajectories for the system, we are now dealing with a structure allowing for more stable points placed in the allowed region than in the previous case. Here, the points $A_{\pm}, E_{\pm}$ lie outside of the physical region. Points $B_{\pm,\pm}$ are all stable. Point $D$ is a saddle point, and is thus unstable. Lastly, the points $C_{-}$ and $C_{+}$ are sources.

Let us notice that any trajectory starting close to the hyperbola $x_1^2 - x_2^2 = 1$ will have $\Omega_\phi \approx 1$, while near the lines $x_1 \pm x_2 = 0$ one obtains $\Omega_\phi\approx 0$, i.e., a universe with a dominating matter component. Thus, if the initial state was sufficiently close to either of the points C$_{\pm}$, one would start with a dark energy-dominated universe, which could then transit to a matter-dominated one, ending up at either of the points $B_{\pm,\pm}$, where the dark energy dominates again.

\begin{figure}[h]
\includegraphics[width=8cm]{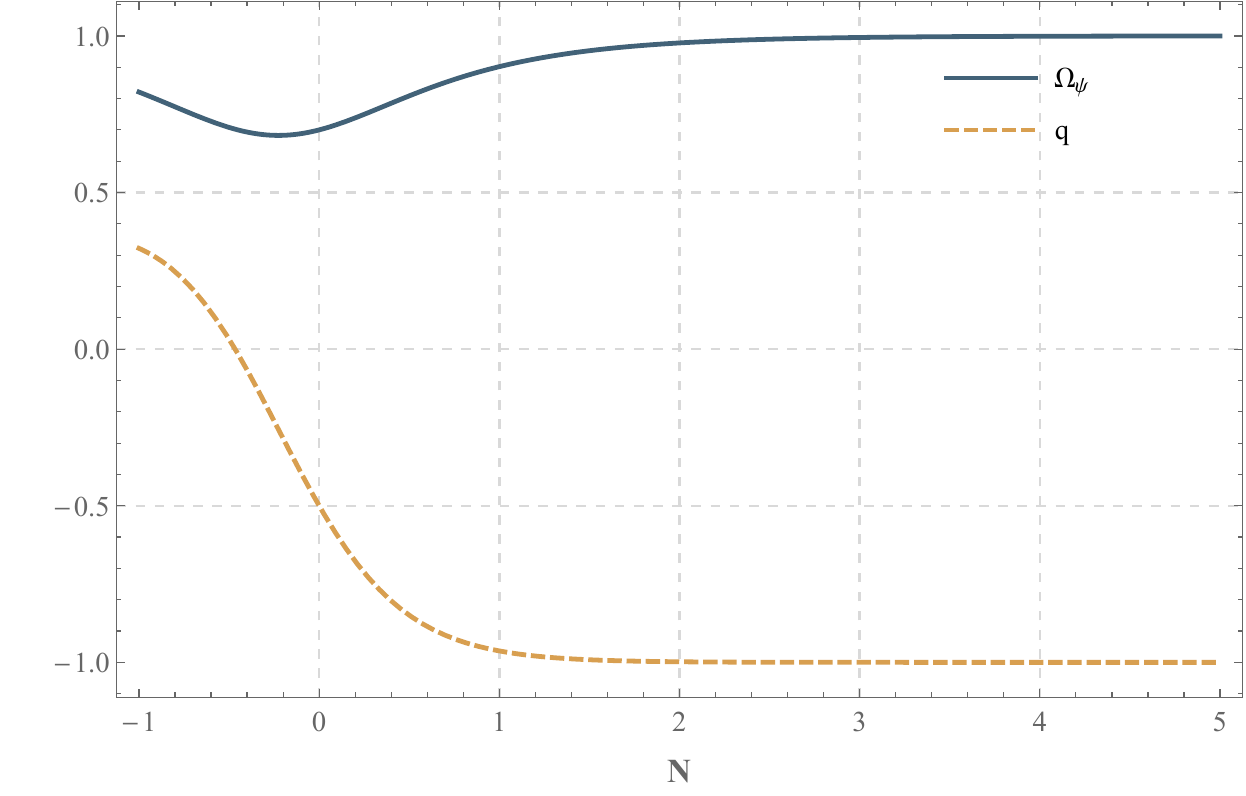}
\caption{Evolution of the deceleration parameter $q$ and the dark energy dimensionless density fraction $\Omega_\phi$ for a trajectory that passes through the point $P$ in Fig \ref{fig:both} at present, i.e. for $\hat{N} = 0$.}
\label{fig:evolution}
\centering
\end{figure}

At the points $B_{\pm,\pm}$, the deceleration parameter becomes $q = -1$, which results in the scale factor at these points $\hat{a}(t) \sim e^{\lambda t}$, with $\lambda$ being some constant, meaning a de Sitter phase. On the other hand, at points $C_{\pm}$ the Hubble parameter blows up, i.e., $\hat{H}\rightarrow \infty$ as we go back in time. Along the trajectories beginning at $C_{\pm}$, the deceleration parameter decreases; see Fig \ref{fig:evolution}, meaning that the cosmic expansion speeds up until the state resembling a universe dominated by the cosmological constant is reached. If we use the fact that current measurements suggest $q \approx -0.55$ \cite{camarena:2020} and $\Omega_\phi \approx 0.7$ \cite{anghanim:2020}, we will be able to set the values of the $x_1, x_2$ parameters at present; in Fig \ref{fig:phase_space_2}, corresponding to the case discussed, it is indicated by the point $P$. Alternatively, one can change the nature of the points $B_{\pm,\pm}$ and make them saddle points, which means that the accelerated period is only transient. Indeed, if one sets $\zeta_2 = 4\zeta_1$, then $B_{\pm,\pm}$ become saddle points, while the points $C_{\pm}$ remain sources, and therefore any trajectory approaching the points $B_{\pm,\pm}$ might be interpreted as a universe undergoing a period of temporary accelerated expansion, preceded by some other type of dynamics of the expansion; this situation is shown in Fig \ref{fig:phase_space_3}.

\section{Discussion and conclusions}

In this work we have investigated a Palatini extension of Horndeski's gravity, focusing in particular on the properties of the model under the change of the disformal frame induced by a disformal transformation of the metric, accompanied by a novel transformation of the independent connection. The transformation of the connection was chosen in such a way that it preserved the form of the action functional required to be form-invariant under both of the transformations. As it has been found, such a general action admits a simple solution for the connection, which turns out to be an auxiliary field, just like in the case of Palatini $f(R)$ theories. For this reason, the theory has a natural metric representation on-shell, i.e., there exists a set of variables, linked to the initial metric and the connection via the disformal transformation of them both, such that the new connection turns out to be a Levi-Civita connection of the new metric tensor. As it also turned out, these new variables are invariant under the disformal change of the initial quantities. Any two sets of the metric tensor and the connection that could be linked with the disformal transformation yield exactly the same metric invariant variables. It remains to be shown that the functions of the scalar field present in the action \eqref{eq:inv_act_ein} are also invariant.

In the second part of the paper, we investigated various cosmological models in the context of the proposed theory in the metric frame. Because the general theory can be very complicated, we decided to make a series of simplifying assumptions, allowing us to obtain particularly simple field equations. It was established that the Palatini extension of Horndeski's gravity under certain assumptions can be thought of as a $k$-essence with a  non-trivial coupling between the scalar field and the matter sector. This coupling introduces a significant level of complexity to the model, as the presence of the kinetic term $X$ in the matter action means that even if the geometric part of the action does not lead to dynamics of the scalar field, it may arise from the matter sector. Obviously, well-posedness and other mathematical properties of such theories remain an open issue. 

In the context of cosmological models, it was rather difficult to obtain analytic solutions, unless the simple case of the vacuum with no cosmological constant present was considered. In the case of dust included in the model, we carried out the dynamical system analysis for two simple cases, involving power-law-type functions for the kinetic term in the action functional. Independently of the value of the exponent, some of the fixed points were common to both of the cases considered, though their exact nature was impacted by the combination of the parameters present in the model. In general, the proposed theory is able to include various phases of cosmic evolution, even though the models discussed above do not seem to correspond exactly to the actual dynamics of the scale factor. 

At this point, one must also observe that the metric was chosen to be the FRLW metric in the metric, Einstein frame. However, under the disformal change, the metric will not in general preserve its form and will describe some other type of the universe. Obviously, this forces us to ask a question analogous to the one posed in the case of classical metric scalar-tensor theories of gravity: which frame is the physical one? This issue pertains to the present theory, too, since it introduces a transformation of the dynamical variables other than the change of the coordinate frame. In a certain way, the dynamics of the system singles out the Einstein frame as the one characterized by simplicity. On the other hand, however, the coupling between the field and the matter sector might lead to a violation of the Weak Equivalence Principle, which could be detectable. 

In principle, the presented work aimed at extending Palatini-Horndeki's gravity beyond the KGB action, with the condition that the metric counterpart, assuming that the connection is an auxiliary field that can be integrated out, reduces to the metric KGB action, being in agreement with the measurements of the speed of gravitational waves and theoretical investigations concerning their decay. The Palatini approach has the advantage of introducing an independent transformation of the connection, so that a disformal change of the metric would not affect all the quantities built from the former (such as the curvature). An interesting problem would be to find a scalar-tensor theory in the Palatini approach such that, on-shell, it would reduce to the currently surviving subclass of Horndeski's gravity, as well as to define set of transformations of the metric and the connection preserving the form of such action. A different continuation of the presented work would be a more general consideration of metric-affine theories, dropping the assumption of symmetry of the connection. In general, one could also consider theories leading to a dynamical connection. Some of the issues mentioned in the paragraph are left as a subject of future work.

\subsection*{Acknowledgments}

The author is grateful to A. Borowiec for his comments and insights, and to G. Leon for discussing the issues related to dynamical system analysis. The author acknowledges financial support from Fondecyt de Postdoctorado 2025, no. 3250036.

\appendix
\section{Scalar field functions in the dynamically-equivalent metric representation}\label{sec:app1}

In this part, we present the transformation formulae for the scalar field functions. Let us notice that, upon expressing the dependence of the 'old' $X$ on the invariant kinetic term $\hat{X}$, we will arrive at formulas written in terms of the scalar field $\phi$ and the term $\hat{X}$. However, it does not mean that the 'new' functions $\hat{B}, \ldots$ are disformally invariant; this issue has not been yet established.
{\footnotesize
\begin{align}
    & \hat{B}(\phi, \hat{X}) = \frac{1}{2 A_1(\phi ,X(\hat{X})) \big(A_1(\phi ,X(\hat{X}))+A_2(\phi
   ,X(\hat{X}))X(\hat{X}) \big)}\Big(A_1(\phi ,X(\hat{X})) \big(X(\hat{X}) C(\phi ,X(\hat{X}))-2 (E_1(\phi ,X(\hat{X}))\nonumber \\
   & \quad\quad\quad\quad  +4
   E_2(\phi ,X(\hat{X}))+X(\hat{X}) E_3(\phi ,X(\hat{X})))\big) +X(\hat{X}) A_2(\phi ,X(\hat{X})) (E_1(\phi,X(\hat{X}))-B(\phi ,X(\hat{X})))\Big), \\
   & \hat{C}(\phi, \hat{X}) = \frac{1}{\left(A_1(\phi, X(\hat{X}))-X(\hat{X}) A_{1, \hat{X}}(\phi, X(\hat{X}))\right)
   (2 A_1(\phi, X(\hat{X}))+3 X(\hat{X}) A_2(\phi, X(\hat{X})))+X(\hat{X})^2 A_1(\phi, X(\hat{X}))
   A_{2, \hat{X}}(\phi, X(\hat{X}))}\nonumber \\
   & \quad\quad\quad\quad \times \frac{1}{(A_1(\phi, X(\hat{X}))+X(\hat{X}) A_2(\phi
   ,X))^{3/2} } \Bigg( 2 \sqrt{A_1(\phi, X(\hat{X}))} \Big(A_1(\phi, X(\hat{X}))^2
   \Big(A_{1, \hat{X}}(\phi, X(\hat{X})) (2 B(\phi, X(\hat{X}))\nonumber\\
   & \quad\quad\quad\quad-5 X(\hat{X}) C(\phi, X(\hat{X}))+4
   E_1(\phi, X(\hat{X}))+16 E_2(\phi, X(\hat{X}))+10 X(\hat{X}) E_3(\phi, X(\hat{X})))\nonumber\\
   & \quad\quad\quad\quad+X(\hat{X})
   A_{2, \hat{X}}(\phi, X(\hat{X})) (-B(\phi, X(\hat{X}))+X(\hat{X}) C(\phi, X(\hat{X}))+4 E_1(\phi
   ,X)+10 E_2(\phi, X(\hat{X}))-2 X(\hat{X}) E_3(\phi, X(\hat{X})))\nonumber \\
   & \quad\quad\quad\quad +A_2(\phi, X(\hat{X})) \Big(X(\hat{X})
   \Big(-5 B_{,\hat{X}}(\phi, X(\hat{X}))+X(\hat{X}) C_{,\hat{X}}(\phi, X(\hat{X}))+6 C(\phi
   ,X)+E_{1, \hat{X}}(\phi, X(\hat{X}))\nonumber \\
  & \quad\quad\quad\quad -8 E_{2, \hat{X}}(\phi, X(\hat{X}))-2 X(\hat{X})
   E_{3, \hat{X}}(\phi, X(\hat{X}))-10 E_3(\phi, X(\hat{X}))\Big)-5 B(\phi, X(\hat{X}))+8
   E_1(\phi, X(\hat{X}))\nonumber \\
   & \quad\quad\quad\quad+10 E_2(\phi, X(\hat{X}))\Big)\Big)+X(\hat{X}) A_1(\phi, X(\hat{X}))
   A_2(\phi, X(\hat{X})) \Big(A_{1, \hat{X}}(\phi, X(\hat{X})) (10 B(\phi, X(\hat{X}))-6 X(\hat{X})
   C(\phi, X(\hat{X}))\nonumber \\
   & \quad\quad\quad\quad-7 E_1(\phi, X(\hat{X}))+6 E_2(\phi, X(\hat{X}))+12 X(\hat{X})
   E_3(\phi, X(\hat{X})))+2 X(\hat{X}) A_{2, \hat{X}}(\phi, X(\hat{X})) (E_1(\phi, X(\hat{X}))-B(\phi
   ,X))\nonumber \\
   & \quad\quad\quad\quad +A_2(\phi, X(\hat{X})) \left(-3 X(\hat{X}) B_{,\hat{X}}(\phi, X(\hat{X}))-6 B(\phi, X(\hat{X}))+3 X(\hat{X})
   E_{1, \hat{X}}(\phi, X(\hat{X}))+6 E_1(\phi, X(\hat{X}))\right)\Big) \nonumber \\
   & \quad\quad\quad\quad+9 X(\hat{X})^2
   A_{1, \hat{X}}(\phi, X(\hat{X})) A_2(\phi, X(\hat{X}))^2 (B(\phi, X(\hat{X}))-E_1(\phi
   ,X))+A_1(\phi, X(\hat{X}))^3 \Big(-2 B_{,\hat{X}}(\phi, X(\hat{X}))\nonumber \\
   & \quad\quad\quad\quad +X(\hat{X}) C_{,\hat{X}}(\phi
   ,X)+5 C(\phi, X(\hat{X}))-2 \left(E_{1, \hat{X}}(\phi, X(\hat{X}))+4
   E_{2, \hat{X}}(\phi, X(\hat{X}))+X(\hat{X}) E_{3, \hat{X}}(\phi, X(\hat{X}))\right)-8
   E_3(\phi, X(\hat{X}))\Big)\Big)\Bigg), \\
   & \hat{F}(\phi, \hat{X}) = \frac{1}{24 A_1(\phi ,X(\hat{X}))^{7/2} (A_1(\phi ,X(\hat{X}))+X(\hat{X}) A_2(\phi
   ,X(\hat{X})))^{3/2}}\Bigg(X(\hat{X}) A_1(\phi ,X(\hat{X}))^2 \Big(4 \Big(-2 E_1(\phi ,X(\hat{X})) \Big(-6
   \Big(A_{1, \phi}(\phi ,X(\hat{X}))\nonumber \\
   & \quad\quad\quad\quad+X(\hat{X}) A_{2, \phi}(\phi ,X(\hat{X}))\Big) +8
   E_2(\phi ,X(\hat{X}))+5 X(\hat{X}) E_3(\phi ,X(\hat{X}))\Big)+E_2(\phi ,X(\hat{X})) \Big(48
   A_{1, \phi}(\phi ,X(\hat{X}))+30 X(\hat{X}) A_{2, \phi}(\phi ,X(\hat{X}))\nonumber \\
   & \quad\quad\quad\quad-4 X(\hat{X}) E_3(\phi
   ,X)\Big)+X(\hat{X}) E_3(\phi ,X(\hat{X})) \Big(12 A_{1, \phi}(\phi ,X(\hat{X}))+3 X(\hat{X})
   A_{2, \phi}(\phi ,X(\hat{X}))+X(\hat{X}) E_3(\phi ,X(\hat{X}))\Big)+3 A_2(\phi ,X(\hat{X}))
   \Big(X(\hat{X})\times \nonumber \\
   & \quad\quad\quad\quad\times\Big(-5 B_{,\phi}(\phi ,X(\hat{X}))+X(\hat{X}) C_{, \phi}(\phi
   ,X)+E_{1, \phi}(\phi ,X(\hat{X}))-8 E_{2, \phi}(\phi ,X(\hat{X}))-2 X
   E_{3,\phi}(\phi ,X(\hat{X}))\Big)+2 F(\phi ,X(\hat{X}))\Big)-11 E_1(\phi
   ,X)^2\nonumber \\
   & \quad\quad\quad\quad+4 E_2(\phi ,X(\hat{X}))^2\Big)+4 B(\phi ,X(\hat{X})) \Big(6
   A_{1, \phi}(\phi ,X(\hat{X}))-3 X(\hat{X}) A_{2, \phi}(\phi ,X(\hat{X}))-4 X(\hat{X}) C(\phi
   ,X(\hat{X}))+14 E_1(\phi ,X(\hat{X}))+20 E_2(\phi ,X(\hat{X}))\nonumber \\
   & \quad\quad\quad\quad+8 X(\hat{X}) E_3(\phi
   ,X(\hat{X}))\Big)-4 X(\hat{X}) C(\phi ,X(\hat{X})) \Big(3 A_{1, \phi}(\phi ,X(\hat{X}))+3 X(\hat{X})
   A_{2, \phi}(\phi ,X(\hat{X}))-5 E_1(\phi ,X(\hat{X}))-2 E_2(\phi ,X(\hat{X}))\nonumber \\
   & \quad\quad\quad\quad+X(\hat{X})
   E_3(\phi ,X(\hat{X}))\Big)-8 B(\phi ,X(\hat{X}))^2+X(\hat{X})^2 C(\phi ,X(\hat{X}))^2\Big)+6 X(\hat{X})^2
   A_1(\phi ,X(\hat{X})) A_2(\phi ,X(\hat{X})) \Big(2 \Big(X(\hat{X}) \Big(E_3(\phi ,X(\hat{X}))\times \nonumber \\
   & \quad\quad\quad\quad
   \Big(A_{1, \phi}(\phi ,X(\hat{X}))+X(\hat{X}) A_{2, \phi}(\phi ,X(\hat{X}))\Big)+3
   A_2(\phi ,X(\hat{X})) \Big(E_{1, \phi}(\phi ,X(\hat{X}))-B_{,\phi}(\phi
   ,X)\Big)\Big)+E_1(\phi ,X(\hat{X})) \Big(-5 A_{1, \phi}(\phi ,X(\hat{X}))\nonumber \\
   & \quad\quad\quad\quad+X(\hat{X})
   A_{2, \phi}(\phi ,X(\hat{X}))-6 E_2(\phi ,X(\hat{X}))-3 X(\hat{X}) E_3(\phi
   ,X)\Big)+E_2(\phi ,X(\hat{X})) \Big(4 A_{1, \phi}(\phi ,X(\hat{X}))+X(\hat{X})
   A_{2, \phi}(\phi ,X(\hat{X}))\Big)\nonumber \\
   & \quad\quad\quad\quad-3 E_1(\phi ,X(\hat{X}))^2\Big)+3 B(\phi ,X(\hat{X}))
   \Big(2 \Big(2 A_{1, \phi}(\phi ,X(\hat{X}))+E_1(\phi ,X(\hat{X}))+2 E_2(\phi
   ,X)+X(\hat{X}) E_3(\phi ,X(\hat{X}))\Big)-X(\hat{X}) C(\phi ,X(\hat{X}))\Big)\nonumber \\
   & \quad\quad\quad\quad+3 X(\hat{X}) C(\phi
   ,X(\hat{X})) E_1(\phi ,X(\hat{X}))\Big)+9 X(\hat{X})^3 A_2(\phi ,X(\hat{X}))^2 \Big((B(\phi
   ,X(\hat{X}))\nonumber \\
   & \quad\quad\quad\quad-E_1(\phi ,X(\hat{X})))^2-4 A_{1, \phi}(\phi ,X(\hat{X})) (-B(\phi ,X(\hat{X}))+2
   E_1(\phi ,X(\hat{X}))+E_2(\phi ,X(\hat{X}))+X(\hat{X}) E_3(\phi ,X(\hat{X})))\Big)+12
   A_1(\phi ,X(\hat{X}))^3 \times \nonumber \\
   & \quad\quad\quad\quad\times\Big(X(\hat{X})\Big(-2 B_{,\phi}(\phi ,X(\hat{X}))+X(\hat{X})
   C_{, \phi}(\phi ,X(\hat{X}))-2 \Big(E_{1, \phi}(\phi ,X(\hat{X}))+4
   E_{2, \phi}(\phi ,X(\hat{X}))+X(\hat{X}) E_{3, \phi}(\phi ,X(\hat{X}))\Big)\Big)\nonumber \\
   & \quad\quad\quad\quad+2 F(\phi
   ,X(\hat{X}))\Big)\Bigg),\\
   & \hat{\Sigma}_1(\phi, \hat{X}) = \frac{\Sigma_1(\phi, X(\hat{X}))}{\sqrt{A_1(\phi, X(\hat{X}))\big(A_1(\phi, X(\hat{X}))+ A_2(\phi, X(\hat{X}))X(\hat{X})\big)}}, \\
   & \hat{\Sigma}_2(\phi, \hat{X}) = \frac{A_2(\phi, X(\hat{X}))}{A_1(\phi, X(\hat{X}))+ A_2(\phi, X(\hat{X}))X(\hat{X})}\Sigma_1(\phi, X(\hat{X}))+\Sigma_2(\phi, X(\hat{X})).
\end{align}}

\end{document}